\documentclass[11pt,a4paper]{article}
\usepackage{amssymb}
\usepackage{graphicx}
\usepackage{hyperref} 
\usepackage[hscale=0.8,vscale=0.8]{geometry}
\renewcommand{\a}{\textsl{a}}

\begin{document}

\title{Quantum Field Theory in Curved Spacetime}
\author{Bernard S.~Kay \\ 
\normalsize Department of Mathematics, 
University of York, York YO10 5DD, U.K.} 

\maketitle

\abstract

The 2023 second edition of a 2006 encyclopedia article on mathematical aspects of quantum field theory in curved spacetimes (QFTCST).  Section-titles (in bold, with new section-titles in bold italics) are: \textbf{Introduction and preliminaries}, \textbf{Construction of a $*$-algebra for a real linear scalar field on globally hyperbolic spacetimes and some general theorems}, \textbf{\textit{More about (quasifree) Hadamard states}}, \textbf{Particle creation and the limitations of the particle concept}, \textbf{Theory of the stress-energy tensor}, \textbf{\textit{More about the intersection of QFTCST with AQFT and the Fewster-Verch No-Go Theorem}}, \textbf{Hawking and Unruh effects}, \textbf{\textit{More about (classical and) quantum fields on black hole backgrounds}}, \textbf{Non-globally hyperbolic spacetimes and the time-machine question}, \textbf{\textit{More about QFT on non-globally hyperbolic spacetimes}}, \textbf{Other related topics and some warnings}.  The article contains many references.  It also includes a review of, and also compares and contrasts, recent results on the implications of QFTCST for the question of the instability of three sorts of Cauchy horizon -- first those inside black holes such as especially Reissner-Nordstr\"om-de Sitter and Kerr-de Sitter,  second the compactly generated Cauchy horizons of spacetimes in which time-machines get manufactured, and third the Cauchy horizon of the spacetime which is believed to describe evaporating black holes and which underlies (one version of) the black hole information-loss puzzle.

\centerline{---}

\noindent
\textit{To the memory of my father, Max Kay (1917-1986) whose love of learning, support, and life-example continue to sustain me.}

\smallskip

\noindent
\textit{And also to Chris Isham, my academic father, together with whom I began the study of this interesting topic 50 years ago!}

\section*{Preface to the Second Edition}

Since the first edition \cite{encyc2006} of this encyclopedia article appeared in 2006,  there has continued to be interesting progress in quantum field theory in curved spacetime (QFTCST) on the mathematical and conceptual side while the subject has continued to play an important r\^ole in many areas of astrophysics and cosmology.   Furthermore QFTCST is relevant (or may soon become relevant) to the analysis of a variety of recent (or recently  proposed) laboratory-based (see e.g.\ \cite{BarCamFuenI,BarCamFuenII}) and space-based (see e.g.\ \cite{KohLouFueBru}) experiments to test aspects of quantum behaviour in gravitational backgrounds and/or in accelerated frames as well as experiments involving condensed matter and fluid etc.\ \emph{analogues} of QFTCST effects such as the Hawking effect \cite{UnruhDumb,BarLibVis,WeinTedPenUnrLaw,RosBerSilLeo,MerMogCheMorBriWez,BrauFaiEtAl}. 
Meanwhile on the theory side, there has recently been renewed interest (see e.g.\ \cite{WittenQFTCST} and see also the final paragraph of this preface) in the lessons that QFTCST may teach us towards the quest for a full quantum theory of gravity.

For all these reasons and more, there continues to be a need for accessible introductions to the subject and I hope that my 2006 encyclopedia article, re-issued with corrections and updates here, may usefully contribute to that purpose as far as the mathematical and conceptual basics of the subject are concerned.  For the more practical and computational aspects of the subject see e.g.\ (DeWitt, 1975) and (Birrell and Davies, 1982) from the `Further Reading' section and/or the  textbooks (Parker and Toms, 2009) and (Mukhanov and Winitzky, 2007) which we have newly added to that section.\footnote{Over the years, there have been many other books, lecture notes, review articles etc.\ with a partly overlapping scope to that of the present article.         See e.g.\ \cite{Gibbons79,Ford97,Jacobson2005,WaldHist12,HollWaldPhysRep,WaldFormul,
BrunFred,BaerGP,FewsterVerch2015,Hack}.    These all differ from the present article and from each other in view of the time at which they were written and in view of their differring emphases on different aspects of the subject.  However all of them contain valuable material and points of view that aren't easily to be found elsewhere.   See also the article \cite{HollandsEncyc} by Hollands and Wald in this encyclopedia.}  

The style of the 2006 edition of the encylopedia demanded that no references be given, but rather a small list of articles and books be appended to each article in a section entitled `Further Reading'.    Aside from a few corrections, additions and updates as well as some small rearrangements of the material, we have preserved the 2006 article, with its list of Further Reading (a total of 18 books and papers to which we have added just two post-2006 textbooks and one collection of articles) to form the core of this second edition.   I believe that this core is still an appropriate vehicle for introducing the subject. 
 
There were a few places in the first edition where we informally specified a reference for some specific result which is not in the Further Reading list by specifying the author name(s) and year of publication, and in those places we have replaced those informal references with full bibliographical references -- included in a new section after the Further Reading section at the end.   These references are indicated in the text by numbers in square brackets.   References to the Further Reading section are instead in the form (Author name(s), year). 

Within that core, we have inserted four new sections (Sections \ref{MoreHad}, \ref{AQFT},  \ref{MoreBH}, \ref{MoreNonGH}) to bring the article fully up to date. One reasonable way to read the article, for someone new to the subject, would be to skip the new sections at a first reading and then return to them (or just to those that seem of interest) later.

We have also added to the bibliography quite a big collection of references to the newly added sections.   Still, as for the 2006 core, the choice of additional material and of additional references inevitably reflects my own interests and tastes and limitations as well as the limitations of space and the bibliography is in no way to be regarded as comprehensive.   Rather, I have attempted to choose a representative selection of articles on the topics I treat which are either self-contained or which have a useful list of references themselves, or are themselves much cited.  I am sure I must have unjustly omitted many important and valuable and relevant references.  But wherever I have done so, I hope the authors will understand that I must have done so inadvertently and forgive me and I would be pleased to be informed about any such omissions.   I would, of course, also be pleased to be informed of any errors.

A feature of the second edition is that it spells out in some detail a number of points which, while somewhat technical, are useful to understand if one wants to contribute to the subject and also fairly easy to explain but which, while likely known to experts, are not so well-known and/or not always easily extractable from the literature.   For example, we include a discussion of the recent correction by Moretti \cite{MorettiCorrn} to the definition in (Kay and Wald, 1991) of the notion of  `global Hadamard' and we also include a brief note on the similarities and differences between the Wald (Wald, 1978) and D\'ecanini-Folacci \cite{DecFol} schemes for the Hadamard renormalization of the stress-energy tensor.  We also provide a simple example that we hope will help the reader to appreciate at least part of the reason why one would expect the Fewster-Verch no-go theorem \cite{FewsterVerch2012a} on the impossibility of any locally covariant preferred state construction to hold.   For similar reasons, in Section \ref{OtherWarn} we have also retained (and added to) our list of warnings.  Aside from that list, Section \ref{OtherWarn} is intended to at least mention a number of topics which, while important, haven't been adequately discussed elsewhere in the article.   Note though that some of the material that was previously in Section \ref{OtherWarn} has now been moved to, and expanded upon, in earlier sections.

One of the most interesting things about QFTCST is that its study has enabled us to address questions about quantum, gravitational, and quantum-gravitational physics which one might have thought would be out of reach of QFTCST because they involve, in an essential way, the quantum and/or dynamical nature of the gravitational field.  Regarding the quantum-gravitational, the impact and influence that our understanding of the Hawking effect has had on string theory and AdS/CFT etc.\ is widely in evidence -- see for example the recent lecture notes \cite{Hartman} of Hartman.  Also one version of the \emph{black hole information-loss puzzle} (see Section \ref{MoreNonGH}) -- which is of course a puzzle about quantum gravity -- has a lot to do with QFT on a particular class of spacetimes -- the \emph{spacetimes of black-hole evaporation} -- that are conjectured to be solutions to semiclassical gravity that correspond to Hawking-evaporating black holes.  As for gravitational physics, a recurring theme in QFTCST concerns the r\^ole quantum effects may have in rendering various sorts of Cauchy horizon impenetrable (or not).  This is reflected in a theme running through the present article -- see our discussion of work in QFTCST related to the question of strong cosmic censorship for Reissner-Nordstr\"om-de Sitter and Kerr-de Sitter Cauchy horizons in Section \ref{MoreBH}; our discussion of time machine-like Cauchy horizons in Section \ref{NonGH}; and, thirdly, our discussion in Section \ref{MoreNonGH} of Ju\'arez-Aubry's recent conjectures \cite{BeniConj} about the Cauchy horizon of a class of spacetimes which includes the spacetimes of black-hole evaporation.  In the course of discussing QFT on all three of these sorts of spacetime with Cauchy horizon we shall also point out some interesting  similiarities and differences between the three cases.

\section{\label{intro} Introduction and Preliminaries}

Quantum Field Theory (QFT) in Curved Spacetime is a hybrid approximate
theory in which quantum matter fields are assumed to propagate in a
fixed classical background gravitational field.  Its basic physical
prediction is that strong gravitational fields can polarize the vacuum
and, when time-dependent, lead to pair creation just as a strong and/or
time-dependent electromagnetic field can polarize the vacuum and/or give
rise to pair-creation of charged particles.  One expects it to be a good
approximation to full quantum gravity provided the typical frequencies
of the gravitational background are very much less than the Planck
frequency ($c^5/G\hbar)^{1/2}\sim 10^{43}s^{-1}$) and provided, with a
suitable measure for energy, the energy of created particles is very
much less than the energy of the background gravitational field or of
its matter sources. Undoubtedly the most important prediction of the
theory is the Hawking effect (Hawking, 1975), according to which a classical black hole will emit thermal radiation at the Hawking temperature $\kappa/2\pi$ where $\kappa$ is the black hole's surface gravity.  E.g.\ in the case of an uncharged nonrotating black hole, $T = (8\pi M)^{-1}$ where $M$ is the black hole's mass. (Here and from now on, we use Planck units  where $G$, $c$, $\hbar$ and $k$ [Boltzmann's constant] are all taken to be 1.)

On the mathematical side, the need to formulate the laws and derive the
general properties of QFT on non-flat spacetimes forces one to state
and prove results in local terms and, as a byproduct, thereby leads to
an improved perspective on flat-space-time QFT too.  It's also
interesting  to formulate QFT on idealized spacetimes with particular
global geometrical features.   Thus, QFT on spacetimes with bifurcate
Killing horizons is intimately related  to the Hawking effect; QFT on
spacetimes with closed timelike curves is intimately related to the
question whether the laws of physics permit the manufacture of a
time-machine.

As is standard in General Relativity, a curved spacetime, $({\cal M}, g)$, is modelled
mathematically as a  (paracompact, Hausdorff) manifold $\cal M$ equipped
with a pseudo-Riemannian metric $g$ of signature $(-,+++)$ (we follow
the conventions of the standard text \cite{MTW}.   We shall also
assume, except where otherwise stated, our spacetime to be {\it globally
hyperbolic}, i.e.\ that $\cal M$ admits a {\it global time coordinate},
by which we mean a global coordinate $t$ such that each constant-$t$
surface is a smooth Cauchy surface  i.e.\ a smooth spacelike 3-surface
cut exactly once by each inextendible causal curve.\footnote{For the equivalence of this statement with other definitions of global hyperbolicity, see \cite{BerSan1}.} (Without this
default assumption, extra problems arise for QFT which we shall discuss, in connection with the time-machine question and the spacetime of black hole evaporation in Sections \ref{NonGH} and \ref{MoreNonGH}.)  
In view of this definition, globally hyperbolic spacetimes are clearly
time-orientable and we shall assume a choice of time-orientation has
been made  so we can talk about the ``future'' and ``past'' directions.
Modern formulations of the subject take, as the fundamental mathematical
structure modelling the quantum field, a $*$-algebra $\cal A$ (with
identity $I$) together with a family of local sub $*$-algebras ${\cal
A}({\cal O})$ labelled by bounded open regions $\cal O$ of the spacetime manifold, $\cal M$,
and satisfying the {\it isotony} or {\it net} condition
that ${\cal O}_1 \subset {\cal O}_2$ implies ${\cal A}({\cal O}_1)$ is a
subalgebra of  ${\cal A}({\cal O}_2)$, as well as the condition that, whenever
${\cal O}_1$ and ${\cal O}_2$ are spacelike separated, then ${\cal
A}({\cal O}_1)$ and ${\cal A}({\cal O}_2)$ commute.   Observables are modelled as algebra elements, $A$, for which $A^*=A$.

Standard concepts and techniques from algebraic quantum theory (AQFT) \footnote{For references on AQFT, see the book (Haag, 2012), the recent review \cite{FewRej} by Fewster and Rejzner and also the article \cite{BuchFred} by Buchholz and Fredenhagen in this encyclopedia.} are then
applicable:  In particular, {\it states} are defined to be positive
(this means $\omega(A^*A) \ge 0$ $\forall A  \in {\cal A}$) normalized
(this means $\omega(I)=1$) linear functionals on $\cal A$.    For an observable $A$, $\omega(A)$ is then interpreted as the expectation value of $A$ in the state $\omega$.   One
distinguishes between {\it pure} states and {\it mixed} states, only the
latter being writeable as non-trivial convex  combinations of other
states. To each state, $\omega$, the {\it GNS-construction} associates a
representation, $\rho_\omega$,  of $\cal A$ on a Hilbert space ${\cal
H}_\omega$ together with a cyclic vector  $\Omega\in{\cal H}_\omega$
such that
$$\omega (A)=\langle\Omega|\rho_\omega(A)\Omega\rangle$$
(and the {\it GNS triple} $(\rho_\omega,{\cal H}_\omega,\Omega)$ is unique up
to equivalence). There are often technical advantages in formulating
things so that the $*$-algebra is a C$^*$-algebra.  Then the GNS
representation is as everywhere-defined bounded operators and  is
irreducible if and only if the state is pure.  A useful concept, due to
Haag, is the {\it folium} of a given state $\omega$ which may be defined
to be the set of all states $\omega_\sigma$ which arise in the form
$\rm{Tr}(\sigma\rho_\omega(\cdot))$ where $\sigma$ ranges over the
density operators (trace-class operators with unit trace) on ${\cal
H}_\omega$.

Given a state, $\omega$, and an automorphism, $\alpha$, which preserves
the state (i.e.\ $\omega\circ\alpha=\omega$) then there will be a unitary
operator, $U$, on ${\cal H}_\omega$ which {\it implements} $\alpha$ in the
sense that  $\rho_\omega(\alpha(A))=U^{-1}\rho_\omega(A)U$ and $U$ is
chosen uniquely by the condition $U\Omega=\Omega$.

On a {\it stationary} spacetime, i.e.\ one which admits a one-parameter
group of isometries whose integral curves are everywhere timelike,  the
algebra will inherit a one-parameter group (i.e.\ satisfying
$\alpha(t_1)\circ\alpha(t_2)=\alpha(t_1+t_2)$) of time-translation
automorphisms, $\alpha(t)$, and, given any stationary state (i.e.\ one
which satisfies $\omega\circ\alpha(t)=\omega\quad \forall t\in {\mathbb R}$)
these will be implemented by a one-parameter group of unitaries, $U(t)$,
on its GNS Hilbert space satsifying $U(t)\Omega=\Omega$.  If $U(t)$ is
strongly continuous so that it takes the form $e^{-iHt}$ and if the
Hamiltonian, $H$, is positive, then $\omega$ is said to be a {\it ground
state}.  Typically one expects ground states to exist and often be
unique.

Another important class of stationary states for the algebra of a
stationary spacetime is the class of {\it KMS states}, $\omega$,
at inverse temperature $\beta$; these have the physical interpretation
of thermal equilibrium states. In the GNS representation of one of
these,  the automorphisms are also implemented by a strongly-continuous
unitary group, $e^{-iHt}$, which preserves $\Omega$ but (in place of $H$
positive) there is a complex conjugation, $J$, on ${\cal H}_\omega$ such that
\begin{equation} 
\label{kms}
e^{-\beta H/2}\rho_\omega(A)\Omega=J\rho_\omega(A^*)\Omega
\end{equation}
for all $A\in {\cal A}$.

An attractive aspect of QFTCST is that its main qualitative
features are already present for linear field theories and, unusually in
comparison with other (non-linear) QFT, these are susceptible of a
straightforward explicit and rigorous mathematical formulation.\footnote{\label{QuantCoroll} What lies behind the tractability of linear QFT is the fact that, unlike for nonlinear QFT, when one formulates the theory as a dynamical system, the quantum path is the same as the classical path.   For Equation (\ref{kg}), this is formalized, as explained in Section \ref{particle},  by Equation (\ref{quantclass}).  Moreover, there appears to be a metatheorem:    \emph{For linear field theories, every quantum property arises as a corollary to some classical property}.  For example, the Reeh-Schlieder property (say) for a wedge (see \cite{KayDoubleWedge}) may be viewed as inherited from a `pre-Reeh-Schlieder' theorem which refers to the one-particle sector of the theory and can be understood as a property of the classical equation (\ref{kg})   (See also Footnote \ref{K}.)   The reader is invited to supply other illustrations.}
In fact, as our principal example, we give in Section (\ref{staralgHad}) a construction for
the field algebra for the quantized real linear Klein Gordon equation
\begin{equation} 
(\Box_g-m^2)\phi=0 \label{kg}
\end{equation} 
of mass $m$ on a globally hyperbolic spacetime $({\cal M},g)$. Here, $\Box_g$
denotes the Laplace Beltrami operator $g^{ab}\nabla_a\partial_b$
($=(|\det(g)|)^{-1/2}\partial_a(|\det(g)|^{1/2}g^{ab}\partial_b)$).  Note that we may, e.g.\ replace $-m^2$ by
$-m^2 - V$ where $V$ is a scalar external background classical field which we will refer to below as a \emph{potential}.  In case $m$ is
zero, taking $V$ to equal $R/6$, where $R$ denotes the Riemann scalar, then 
makes the equation {\it conformally invariant} and the equation is then called the
{\it conformally coupled} Klein Gordon equation.  In the case of a vacuum solution to Einstein's equations, $R$ of course vanishes and so the minimally coupled and conformally coupled equations are the same.  However, the stress-energy tensors, $T_{ab}$, defined as $-2|\det(g)|^{-{1/2}}\delta/\delta g_{ab}S$ -- where the action, $S$, is the integral with the volume element $|\det(g)|^{1/2}d^4x$ of $L$ where $L$ is, respectively, $-\frac{1}{2}(g^{ab}\partial_a\phi\partial_b\phi + m^2\phi^2)$ and $-\frac{1}{2}(g^{ab}\partial_a\phi\partial_b\phi + [m^2+ \frac{1}{6}R]\phi^2$) -- will, of course, differ.

The main new feature of quantum field theory in curved spacetime
(present already for linear field theories) is that, in a general
(neither flat, nor stationary) spacetime there will not be any single
preferred state but rather a family of preferred states, members of
which are best regarded  as on an equal footing with one-another. It is
this feature which makes the above algebraic framework particularly
suitable,  indeed essential, to a clear formulation of the subject.
Conceptually it is this feature which takes the most getting used to.
In particular, one must realize that, as we shall explain in Section \ref{particle},
the interpretation of a state as having a particular
``particle-content'' is in general problematic because it can only be
relative to a particular choice of ``vacuum'' state and, depending on
the spacetime of interest, there may be one state or several states  or,
frequently, no states at all which deserve the name ``vacuum'' and even
when  there are states which deserve this name, they will often only be
defined in some approximate or asymptotic or transient sense or only on
some subregion of the spacetime.

Concomitantly, one does not expect global observables such as the
``particle number'' or the quantum Hamiltonian of flat-spacetime free
field theory to generalize to a general curved spacetime context, and for this
reason local observables play an essential r\^ole in the theory.  The
(quantized renormalized) stress-energy tensor, $T_{ab}$, is a particularly natural and important
such local observable and the theory of this is central to the whole
subject.  A brief introduction to it is given in Section \ref{stress}.

The main reason for our interest in the quantized stress energy tensor is that its expectation value, $\omega(T_{ab})$, in a quantum state, $\omega$, will serve as the right hand side of the \emph{semiclassical Einstein equations} $G_{ab} = 8\pi G_N \omega(T_{ab})$ which, when solved simultaneously with a suitable QFTCST equation such as (\ref{kg}), is expected, in some situations, to give us some sort of approximate indication of what the predictions of quantum gravity would be while staying within the framework of a classical, but now dynamical, spacetime.

There is then a new section, \ref{AQFT}, describing some recent developments in AQFT, focusing on the point of view of \emph{locally covariant QFT} and discussing the Fewster-Verch no-go theorem on the non-existence of a locally covariant preferred state.  Section \ref{blackholes} is on the Hawking and Unruh effects.  Section \ref{MoreBH} has further information about black hole backgrounds including a discussion of classical and quantum scattering theory on such backgrounds. Section \ref{NonGH} is on the problems of extending the theory beyond the ``default'' setting,  to non-globally hyperbolic spacetimes. Here, we focus on the question of whether it is possible in principle to manufacture a time-machine, and, in a new section, Section \ref{MoreBH}, we also, discuss QFT on the (non-globally hyperbolic) spacetime which is believed to give an approximate description of black-hole evaporation and which underlies one version of the black-hole information loss puzzle.   Finally, Section \ref{OtherWarn}  briefly mentions a number of other interesting and active areas of the subject as well as issuing a few {\it warnings} to be borne
in mind when reading the literature.  

\section{\label{staralgHad} Construction of a $*$-Algebra for a Real Linear Scalar Field on
Globally Hyperbolic Spacetimes and Some General Theorems}

On a globally hyperbolic spacetime, the classical equation (\ref{kg})
admits well-defined {\it advanced} and {\it retarded Green functions}
(strictly bidistributions) $\Delta^A$ and $\Delta^R$ and the standard
covariant quantum free real (or ``Hermitian'') scalar field commutation
relations familiar from Minkowski spacetime free field theory naturally
generalize to the (heuristic) equation
$$[\hat\phi(x), \hat\phi(y)]=i\Delta(x,y)I$$
where $\Delta$ is the {\it Lichnerowicz commutator function} (again, strictly bidistribution)
$\Delta=\Delta^A-\Delta^R$.  Here, the ``$\hat{\quad}$'' on the quantum
field $\hat\phi$ serves to distinguish it from a classical solution
$\phi$. In mathematical work, one does not assign a meaning to the field
at a point itself, but rather aims to assign meaning to {\it smeared
fields} $\hat\phi(F)$ for all real test functions $F\in
C_0^\infty({\cal M})$\footnote{Here and elsewhere, when we write $C_0^\infty(X)$ for some manifold $X$, we mean, by default, $C_0^\infty(X, \mathbb{R})$.}  which are then to be interpreted as standing for
$\int_{\cal M}\hat\phi(x)F(x)|\det(g)|^{1/2}d^4x$. In fact, it is
straightforward to define a {\it minimal field algebra} (see below)
${\cal A}_{\mathrm{min}}$ generated by such $\hat\phi(F)$ which satisfy
the suitably smeared version
$$[\hat\phi(F_1), \hat\phi(F_2)]=i\Delta(F_1,F_2)I$$
of the above commutation relations together with  Hermiticity (i.e.\
$\hat\phi(F)^*=\hat\phi(F)$), the property of being a weak solution of
the equation (\ref{kg}) (i.e.\   $\hat\phi((\Box_g-m^2)F)=0\quad
\forall F\in C_0^\infty({\cal M})$) and linearity in test functions.
There is a technically different alternative formulation of this minimal
algebra, which is known as the {\it Weyl algebra}, which is constructed
to be the C$^*$ algebra generated by operators $W(F)$ (to be interpreted
as standing for $\exp(i\int_{\cal M}\hat\phi(x)F(x)|\det(g)|^{1/2}d^4x$) satisfying
$$W(F_1)W(F_2)=\exp(-i\Delta(F_1,F_2)/2)W(F_1+F_2)$$
together with $W(F)^*=W(-F)$ and $W((\Box_g-m^2)F)=I$.\footnote{If $(S, \sigma)$ denotes the real vector space, $S$, of classical real-valued smooth solutions of (\ref{kg}) with compact support on Cauchy surfaces equipped with the symplectic form $\sigma$ defined such that 
$\sigma(\phi_1,\phi_2)$ is equal to the $\sigma((f^1_0,p^1_0);(f^2_0,p^2_0))$ of Equation (\ref{sigma}) in Section \ref{particle} where $(f^1,p^1)$, $(f^2,p^2)$ are the Cauchy data of $\phi_1, \phi_2$ on any Cauchy surface (which we identify here with the ${\cal C}_0$ of Section \ref{particle}) then what is usually called the \emph{Weyl algebra over $(S, \sigma)$} is the C$^*$ algebra generated by operators $\tilde W(\phi)$ (to be interpreted
as standing for $\exp(i\sigma(\hat\phi, \phi)$) satisfying
$\tilde W(\phi_1)\tilde W(\phi_2)=\exp(-i\sigma(\phi_1,\phi_2)/2)\tilde W(\phi_1+\phi_2)$ and the operator $W(F)$ of the main text may be identified with $\tilde W(\phi)$ where $\phi$ is equal to the solution $\Delta F$ (where the distribution $\Delta F$ is defined by $(\Delta F)(F_1) = \Delta(F_1,F)$ for all test functions $F_1 \in C_0^\infty(\cal M)$.).   See Section 3 of (Kay and Wald, 1991).}    With either
the minimal algebra or the Weyl algebra one can define, for each bounded
open region $\cal O$, subalgebras ${\cal A}({\cal O})$ as generated by the
$\hat\phi(\cdot)$ (or the $W(\cdot)$) smeared with test functions
supported in ${\cal O}$ and verify that they satisfy the above ``net''
condition and commutativity at spacelike separation.

Specifying a state, $\omega$, on ${\cal A}_{\mathrm{min}}$  is tantamount
to specifying its collection of {$n$-point distributions} (i.e.\ smeared
$n$-point functions) $\omega(\hat\phi(F_1)\dots\hat\phi(F_n))$.\footnote{\label{WeylReg} In the
case of the Weyl algebra, one restricts attention to ``regular'' states
for which the map $F\rightarrow \omega(W(F))$ is sufficiently often
differentiable on finite dimensional subspaces of $C_0^\infty({\cal M})$
and defines the $n$-point  distributions in terms of derivatives with
respect to suitable parameters of expectation values of suitable Weyl 
algebra elements.   For example, one may define the anticommutator function, $G(F_1,F_2)$, of a state $\omega$, (in place of (\ref{anticom})) to be $-2\partial^2W(t_1F_1 + t_2F_2)/\partial t_1\partial t_2|_{t_1=t_2=0}$ and the smeared two-point function, $\omega(\phi(F_1)\phi(F_2))$, will then be $(G(F_1,F_2) + i\Delta(F_1,F_2))/2$.}  A particular r\^ole is played in the theory by the
{\it quasi-free} states for which all the {\it truncated} (see (Haag, 2012)) $n$-point
distributions except for $n=2$ vanish.  Thus all the $n$-point
distributions for odd $n$ vanish while the 4-point distribution is made
out of the 2-point distribution according to
$$\omega(\hat\phi(F_1)\hat\phi(F_2)\hat\phi(F_3)\hat\phi(F_4))=
\omega(\hat\phi(F_1)\hat\phi(F_2))\omega(\hat\phi(F_3)\hat\phi(F_4))$$
$$+\omega(\hat\phi(F_1)\hat\phi(F_3))\omega(\hat\phi(F_2)\hat\phi(F_4))
+ \omega(\hat\phi(F_1)\hat\phi(F_4))\omega(\hat\phi(F_2)\hat\phi(F_3))$$
etc.\footnote{\label{K} Each quasi-free state, $\omega$, can be thought of as arising from a \emph{one-particle Hilbert space structure} $(K, \cal H)$ over $(S, \sigma)$.  This consists of a \emph{one-particle Hilbert space} $\cal H$ and a real-linear map, $K$ from $S$ to $\cal H$ such that the complexified range of $K$ is dense in $\cal H$ and $K$ is symplectic in the sense that $\forall \phi_1, \phi_2\in S$,  $2\Im\langle K\phi_1|K\phi_2\rangle = \sigma(\phi_1, \phi_2)$.   The anticommutator function $G(F_1, F_2)$ of (\ref{anticom}) is then equal to $2\langle K\Delta F_1|K\Delta F_2\rangle$ and in fact, in a Weyl algebra formulation, we have $\omega(W(F)) = \exp(-\|KF\|^2/2)$ where $\|\cdot\|$ denotes the norm in the one-particle Hilbert space.  Also in this Weyl algebra case, $\omega$ is pure if and only if the range of $K$ is dense (without the need for complexification).  For the proof of this and more information about one-particle Hilbert-space structures, see Appendix A in (Kay and Wald, 1991).   When the underlying spacetime is stationary, the space of solutions, $(S,\sigma)$, will be equipped with a one-parameter group, ${\cal T}(t)$ of symplectomorphisms and one can seek a \emph{one-particle structure} (in the sense introduced in \cite{KayThesis}) consisting of a one-particle Hilbert space structure $(K, \cal H)$ (with $\mathrm{ran}K$ dense) for $(S,\sigma)$ together with a unitary group, $e^{-iHt}$ for a positive Hamiltonian, $H$, such that $U(t)K = K{\cal T}(t)$.   The corresponding state $\omega$ is then a ground state.  (See \cite{KaySeg} for the uniqueness of such one-particle structures and \cite{KayCMP78} for constructions for a wide class of stationary spacetimes.)   There is also a notion \cite{KayKMS} of `KMS one-particle structure' for which the corresponding state, $\omega$, is a KMS state.)}  The anticommutator distribution
\begin{equation} 
\label{anticom}
G(F_1,F_2) =
\omega(\hat\phi(F_1)\hat\phi(F_2))+\omega(\hat\phi(F_2)\hat\phi(F_1))
\end{equation}
of a quasi-free state (or indeed of any state) will satisfy the properties
(for all test functions $F$, $F_1$, $F_2$ etc.):

\begin{description}

\item{(a)} {\it (symmetry)} $G(F_1,F_2)=G(F_2,F_1)$

\item{(b)} {\it (weak bisolution property)}
$$G((\Box_g - m^2)F_1,F_2)=0=G(F_1,(\Box_g - m^2)F_2)$$

\item{(c)} {\it (positivity)} $G(F,F)\ge0$ and
$G(F_1,F_1)^{1/2}G(F_2,F_2)^{1/2}\ge |\Delta(F_1,F_2)|$

\end{description}

\noindent
and it can be shown that, to every bilinear functional $G$ on
$C_0^\infty({\cal M})$ satisfying (a), (b) and (c), there is a
quasi-free state with two-point distribution $(1/2)(G + i\Delta)$. 
One further declares a quasi-free state to be {\it physically admissible}
only if (for pairs of points in sufficiently small convex
neighborhoods) it satisfies the

\begin{description}

\item{(d)} {\it ([local] Hadamard condition)}   The unsmeared anticommutator function, $G(x_1,x_2)$, takes the form  ``$G(x_1,x_2)={1\over 2\pi^2}\left
(u(x_1,x_2){\rm P}{1\over\sigma}+v(x_1,x_2)\log|\sigma|+w(x_1,x_2)\right )$''\footnote{This formula implicitly involves a choice of length scale, say $L$ and the middle term should more strictly be written $v(x_1,x_2)\log(|\sigma|/L)$.   Changing the length scale from $L$ to, say, $L'$ will then be accompanied by the change from $w$ to $w + v\log(L/L')$.}
where $\sigma$ denotes the
(signed) geodetic interval between $x_1$ and $x_2$. $u$ (which
satisfies $u(x_1,x_2)=1$ when $x_1=x_2$) and $v$ are certain smooth
two-point functions determined in terms of the local geometry (and the
local values of any potential term $V$) by something called the {\it Hadamard procedure}
while the smooth two-point function $w$ depends on the state.

\end{description}

\noindent
This last condition expresses the requirement that (locally) the
two-point distribution actually ``is'' (in the usual sense in which one
says that a distribution ``is'' a function) a smooth function for pairs
of non-null-separated points, and at the same time requires that the
two-point distribution be singular at pairs of null-separated points and
locally specifies the nature of the singularity for such pairs of points
with a leading ``principal part of $1/\sigma$'' type singularity and a
subleading ``$\log|\sigma|$'' singularity.  The  important point is that this local Hadamard condition
on the two-point distribution is believed to be the correct
generalization to a curved spacetime of the well-known universal
short-distance behaviour  shared by the (truncated) two-point
distributions of all physically relevant  states for the special case of
our theory when the spacetime is flat (and, in case one adds a $V$ to $m^2$, $V$ vanishes).  In the latter
case, $u$ reduces to $1$, and $v$ can be expressed as a simple power series
$\sum_{n=0}^\infty v_n\sigma^n$ with $v_0=m^2/4$ etc.  (In the general case, $u$ is the square root of the \emph{Van Vleck-Morette determinant}  and one again has a power series $\sum_{n=0}^\infty v_n\sigma^n$ for $v$ but where the coefficients, $v_n$, are, in general, symmetric functions of $x_1$ and $x_2$ which are determined by certain \emph{Hadamard recursion relations}.)

Actually, it is known (this is the content of ``Kay's Conjecture'' which
was proved \cite{RadzKayConj} by M. Radzikowski in 1992) that (a), (b), (c) and (d)
together imply that the two-point distribution is nonsingular at all
pairs of (not necessarily close-together) spacelike-separated points.
More important than this result itself is a replacement for the
Hadamard condition which Radzikowski \cite{RadzMicro} originally introduced as a tool towards its proof:

\begin{description}

\item{(d$'$)} {\it (Wave Front Set [or Microlocal] Spectrum Condition)}
WF$(G+i\Delta)= \lbrace (x_1,p_1;x_2,p_2)\in
T^*({\cal M}\times {\cal M})\diagdown{\bf 0}| \,
x_1$ and $x_2$ lie on a single null geodesic, $p_1$ is cotangent to
that null geodesic and future pointing, and $p_2$ when parallel
transported along that null geodesic from $x_2$ to $x_1$ equals
$-p_1\rbrace$

\end{description}

\noindent
Here, the notion of the {\it wave front set of a distribution}, $D$, or WF$(D)$ for short, is a notion 
from {\it microlocal analysis}.
For the gist of what it means, it suffices to know that to say that an
element $(x,p)$ of the cotangent bundle of a manifold (excluding the
zero section ${\bf 0}$) is in the {\it wave front set}, WF, of a given
distribution, $D$, on that manifold may be expressed informally by saying that
that $D$ is singular {\it at} the point $x$ {\it in the
direction} $p$.  (And here the notion is applied in the case that $D$ is $G+i\Delta$, thought
of as a distribution on the manifold ${\cal M}\times {\cal M}$.)

For many questions, the microlocal spectrum condition, (d$'$) above, is a more convenient and powerful technical starting point than the original formulation, (d), of the Hadamard condition.  Early applications of (d$'$) were to the question of time-machines (Kay, Radzikowski and Wald, 1997) -- see Section \ref{NonGH} -- and to certain questions about energy inequalities (see e.g.\ \cite{FewWorLin2000}) and Wick powers (see \cite{FewVerNecess} and references therein), while it is an important ingredient of pAQFT -- see Section \ref{OtherWarn}.  However, it seems that the original formulation, (d), cannot be dispensed with altogether.   Thus it continues to be needed to define the Hadamard renormalization scheme -- see Section \ref{stress} -- for the stress-energy tensor, and there are questions, including questions involving the restriction of two-point functions to (spacelike and lightlike) surfaces, where (cf.\ the remarks at the end of Section 4.2.1 in \cite{KoSan2013}) (d) still seems to be needed.  For example, in its `global' version discussed in Section \ref{MoreHad}, (d) is essential for the analysis of stationary Hadamard states on spacetimes with bifurcate Killing horizons in (Kay and Wald, 1991) and of QFT near other types of horizons -- see, for example, \cite{KurpiczEtAl}.

We remark that generically (and, e.g., always if the spatial sections are
compact and, in case one adds a (constant in time) potential term $V$, if $m^2 + V(x)$ is everywhere positive) the Weyl algebra for equation (\ref{kg})  on a given stationary spacetime will have a unique ground state and unique KMS states at each temperature and these will be quasi-free.   Also \cite{SahlVerch} (see also \cite{KoSan2013}) it is known that ground states and KMS states for (\ref{kg}) and many other linear field theories on stationary spacetimes will satisfy (d$'$) (or the generalizations thereof to other spins etc.) -- i.e.\ will be Hadamard.

Quasi-free states are important for the theory of Equation (\ref{kg}) also because of a theorem of Verch \cite{VerchConjProof}
(in verification of another conjecture of Kay) that (in the Weyl
algebra framework) on the algebra of any bounded open region, the folia
of the quasi-free Hadamard states coincide. With this result one can
extend the notion of physical admissibility to
not-necessarily-quasi-free states by demanding that, to be admissible, a
state belong to the resulting common folium when restricted to the
algebra of each bounded open region;  equivalently that it be a {\it
locally normal} state on the resulting natural extension of the net of
local Weyl algebras to a net of local W$^*$ algebras.


As for general (not necessarily quasi-free) states.   We can still define a Hadamard state to be a state whose two-point function satisfies the Hadamard condition.  Ostensibly, this makes no restriction on the state's $n$-point functions for other $n$.   However a result of Sanders \cite{SanHad} entails the important fact that, actually, and essentially thanks to the positivity condition (c), the truncated $n$-point functions of all Hadamard states for $n \ne 2$ are necessarily smooth.


\section{\label{MoreHad} \textit{More about Hadamard states}}

For further reading on quasi-free Hadamard states, see the article \cite{KhavMor} in (Brunetti et al., 2005).

Before the work of Radzikowski mentioned above, a notion of {\it global Hadamard} was defined in Section 3 of (Kay and Wald, 1991) for the quantum field theory of (\ref{kg}) on a general globally hyperbolic spacetime which incorporates both the local Hadamard condition and the absence of nonlocal spacelike singularities.

An important correction to that definition was made \cite{MorettiCorrn} by Moretti in 2021.\footnote{The error in (Kay and Wald, 1991) which Moretti points out is in the sentence: ``The squared geodesic distance, $\sigma$, will be well-defined and smooth in an open neighborhood, $\cal O$, in $M \times M$, of the set of causally related points $(x,y)$ such that $J^+(x)\cap J^-(y)$ and $J^-(x)\cap J^+(y)$ are contained within a convex normal neighborhood.'' The point is that while, for any $(x,y)$  in that set of causally related points, say with $y$ to the future of $x$, a preferred geodesic connecting $x$ and $y$ can be picked out by demanding that it stays inside $J^+(x)\cap J^-(y)$, for spacelike-related $(x,y)$ in the open neighborhood $\cal O$ (which may well be connected by more than one geodesic) there is no preferred convex causal set the staying inside of which could be used as a selection criterion to pick out a preferred exemplar.   And the different such geodesics might well have different lengths.   However, it is shown in \cite{MorettiCorrn} that as long as $\cal M$ is paracompact, the resulting different values for the (singular) first two terms of the Hadamard form in (d) above, will differ by a smooth two-point function.}   The (thus corrected) notion of global Hadamard is initially only defined on a {\it causal normal neighborhood}, $\cal N$, of a Cauchy surface, $\cal C$ -- a notion introduced, and shown to always exist, in Section 2 of (Kay and Wald, 1991) which is defined by the property that every two causally related points in $\cal N$ belong to a common convex normal neighborhood and thus are connected in $\cal N$ by a unique causal geodesic.  One then argues that this global Hadamard property propagates in the sense that the set of states which are Hadamard in a causal normal neighborhood of one Cauchy surface, say ${\cal C}_1$, is the same as the set of states which are Hadamard in a causal normal neighborhood of any other Cauchy surface, ${\cal C}_2$.\footnote{A proof of this was sketched in (Kay and Wald, 1991), a forerunner having been given in \cite{FullSweeWald}.  An alternative proof based on version (d$'$) of the (global) Hadamard condition and the microlocal propagation of singularities theorem was given in \cite{RadzMicro}.   We give in the rest of this footnote a version of the proof in (Kay and Wald, 1991).   Clearly it suffices to prove it for  ${\cal C}_2$, to the future of ${\cal C}_1$.  One can then take an arbitrary compact submanifold, ${\cal S}_2$ of ${\cal C}_2$.  The (trapezium-shaped) region to the past of ${\cal S}_2$ and to the future of ${\cal C}_1$ will (if its boundary is included) itself be compact and therefore coverable with a finite number of overlapping causal normal neighborhoods of a finite number of Cauchy surfaces which can be taken to be constant-time surfaces for a certain finite set of values of a suitable time-coordinate such that ${\cal S}_2$ and the causal past of ${\cal S}_2$ in ${\cal C}_1$ are themselves constant time surfaces.  One may then see that the Hadamard property propagates from the earlier to the later  of any pair of adjacent overlapping such neighborhoods, say ${\cal N}_i$ and ${\cal N}_{i+1}$, by first noting that on their overlap, ${\cal N}_i \cap {\cal N}_{i+1}$, the difference between the two-point function of any global Hadamard state on ${\cal N}_{i+1}$ and any global Hadamard state on ${\cal N}_i$ will be a smooth bisolution and then by looking at the Cauchy evolution of the Cauchy data for this bisolution on a Cauchy surface for the overlap. One can then obviously propagate from the earliest to the latest causal normal neighborhood in a finite number of such steps.  The result follows since the choice of ${\cal S}_2$ was arbitrary.}

The above propagation result may be combined (as explained in \cite{FullNarcoWald}) with a smooth deformation argument to show that, once one has a construction of, or proof of the existence of, a (global) Hadamard state for Equation (\ref{kg}) on some particular globally hyperbolic spacetime with a given topology then there must be many globally Hadamard states on any globally hyperbolic spacetime with the same topology.   For this deformation argument in the case of a massive scalar field, the particular spacetime with a given topology was chosen, in \cite{FullNarcoWald}, to be an ultrastatic spacetime -- i.e.\ a static spacetime where the metric takes the form $-dt^2 + ^{(3)}\!\!g_{ij}dx^idx^j$ -- with the desired topology and it was shown in \cite{FullNarcoWald} (or follows from the more recent more general argument for stationary spacetimes in \cite{SahlVerch}) that ground states on such spacetimes are Hadamard states.

However, this still leaves open many questions as to whether we can construct, and have some control over the properties of, Hadamard states for certain given spacetimes or classes of spacetimes and leaves unanswered questions as to whether certain special states with certain special properties on certain special spacetimes are (also) Hadamard states.   To amplify on this point: (a) the deformation method of \cite{FullNarcoWald} is fairly inexplicit -- it is good for showing that there are many Hadamard states but not for computing with them;
(b) the deformation method would not work for spin-3/2 and spin-2 fields, where the background spacetime has to satisfy the Einstein equations for the appropriate classical stress-energy tensor (to be solved simultaneously with the classical equation(s) of the matter field(s) in the problem), so cannot be smoothly deformed to ultrastatic;
(c) there is a similar problem to show that there are states obeying the analytic wavefront set condition (cf.\ e.g.\ \cite{StrohEtAl2002}) because one cannot make the deformation in an analytic way.\footnote{I thank Chris Fewster for making the points (a), (b), (c).}      One also wants to generalize to field theories other than that of the scalar equation (\ref{kg}). 

Besides ground and KMS states for stationary spacetimes as discussed above, much more work has been done, in aid of answering the many open questions referred to above, on the construction of Hadamard states on general (globally hyperbolic) spacetimes as well as on special classes of spacetimes such as asymptotically flat spacetimes and spacetimes with bifurcate Killing horizons (see Section \ref{blackholes}) using techniques including the theory of pseudo-differential operators \cite{JunkerRev,GerWrochPseud}, the theory of Calder\'on projectors \cite{GerWrochCald}, scattering theory \cite{GerWrochInOut,MoMuVo} and the characteristic Cauchy problem \cite{DapMorPinHadLight,GerWrochChar} both for the covariant Klein Gordon equation as well as the Dirac field \cite{HollDirac,GerStosk} and the Proca and Maxwell and linearized Yang Mills fields \cite{FewPfen,WrochnaZahn}, while G\'erard and Wrochna have also explicitly generalized the deformation method of \cite{FullNarcoWald} to linearized Yang Mills in \cite{GerardWrochnaYM} and also given \cite{GerardMurroWrochnaLinGR} a construction of Hadamard states for linearized quantum gravity -- in the harmonic gauge on analytic backgrounds of bounded geometry.


Let us also mention further work on the Proca field \cite{MomMuVo2} and also \cite{FewPolar} (which includes an important correction to \cite{MomMuVo2}).   The latter work is based on an extension \cite{FewGreen} of the notion of Hadamard states to a large class of linear dynamical equations which are `Green-Hyperbolic'.

There is also a way \cite{BrumFred} (see also \cite{FewArtState}) of obtaining a class of ``vacuum-like'' Hadamard states, known as \emph{Brum-Fredenhagen} (or \textit{BM}) states, by making a certain modification of a certain construction of a preferred state for a general spacetime known as the \emph{Sorkin-Johnston} (or \textit{S-J}) state \cite{SorkJohnst} which however itself fails, in general, to be Hadamard!  Let us also remark here that the S-J state construction also fails to be locally covariant and thus (see \cite{FewVerNecess,FVRecentConstr,FewArtState}) is not in conflict with the Fewster-Verch no-go theorem which we will discuss in Section \ref{AQFT}.

Progress on demonstrating the Hadamard nature of certain states on black hole spacetimes and other spacetimes with bifurcate Killing horizons will be discussed in Section \ref{blackholes}.

It is natural to wonder whether the Hadamard condition can be replaced with some approximation or other.   E.g.\ one might contemplate truncating the power series that defines the Hadamard $v$ function at some finite order.  In this connection, it is worth noting that Fewster and Verch have proven \cite{FewVerNecess} that the Wick squares of all time derivatives of the quantized Klein-Gordon field have finite fluctuations only if the Wick-ordering is defined with respect to an exactly Hadamard state.

Lastly, let us mention that, in the context of semiclassical gravity, a proposal has recently been made \cite{JuaKayMirSud} for a notion of \emph{surface Hadamard} which is conjectured to be a suitable condition on gravitational and quantum initial data on an initial surface to ensure that the Cauchy development of that initial data exists and is a Hadamard solution of semiclassical gravity.   Here again, one might contemplate approximating this notion.   But, as is pointed out in \cite{JuaKayMirSud} (see Footnote 5 there), any formulation of semiclassical gravity that admits states, $\omega$, which are not exactly Hadamard would involve metrics that have a lower degree of smoothness than $C^\infty$.

\section{\label{particle} Particle Creation and the Limitations of the Particle Concept}

Global hyperbolicity also entails that the Cauchy problem is well posed
for the classical field equation (\ref{kg}) in the sense that for every
Cauchy surface, ${\cal C}$, and every pair $(f,p)$ of Cauchy data in
$C_0^\infty({\cal C})$,  there exists a unique solution $\phi$ in
$C_0^\infty({\cal M})$ such that $f=\phi|_{\cal C}$ and
$p=|\det(g)|^{1/2}g^{0b}\partial_b\phi|_{\cal C}$.  Moreover $\phi$ has
compact support on all other Cauchy surfaces.  Given a global time
coordinate $t$, increasing towards the future, foliating $\cal M$ into a
family of contant-$t$ Cauchy surfaces, ${\cal C}_t$, and given a choice
of global time-like vector field $\tau^a$ (for example, for suitable $t$ \cite{BerSan2} one may take
$\tau^a=g^{ab}\partial_b t$)
enabling one to identify all the ${\cal C}_t$, say with ${\cal C}_0$, by identifying points cut by the same
integral curve of $\tau^a$, a single such classical solution $\phi$ may
be pictured as a family $\lbrace (f_t, p_t): t \in {\mathbb R}\rbrace$  of
time-evolving Cauchy data on ${\cal C}_0$.  Moreover, since (\ref{kg})
implies, for each pair of classical solutions, $\phi_1$, $\phi_2$, the
conservation (i.e.\ $\partial_a j^a=0$) of the current
$j^a=|\det(g)|^{1/2}g^{ab}(\phi_1\partial_b\phi_2 -
\phi_2\partial_b\phi_1)$, the symplectic form 
(on $C_0^\infty({\cal C}_0)\times
C_0^\infty({\cal C}_0)$)
\begin{equation}
\label{sigma}
\sigma((f^1_t,p^1_t);(f^2_t,p^2_t))=
\int_{{\cal C}_0}(f^1_tp^2_t-p^1_tf^2_t)d^3x
\end{equation}
will be conserved in time.  ($d^3x$ denotes $dx^1\wedge dx^2\wedge dx^3$.)

Corresponding to this picture of classical dynamics, one expects there
to be a description of quantum dynamics in terms of a family of
sharp-time quantum fields $(\varphi_t, \pi_t)$ on ${\cal C}_0$,
satisfying heuristic canonical commutation relations
$$[\varphi_t({\bf x}),\varphi_t({\bf y})]=0, \quad [\pi_t({\bf x}),
\pi_t({\bf y})]=0, \quad
[\varphi_t({\bf x}), \pi_t({\bf y})]=i\delta^3({\bf x},{\bf y})I$$
and evolving in time according to the same dynamics as the Cauchy data
of a classical solution.  (Both these expectations are justified because
the field  equation is linear.)  An elegant way to make rigorous
mathematical sense of these expectations is in terms of a  $*$-algebra
with identity generated by Hermitian objects ``$\sigma((\varphi_0,
\pi_0); (f,p))$'' (``symplectically smeared sharp-time fields at
$t=0$'') satisfying linearity in $f$ and $p$ together with the
commutation relations
$$[\sigma((\varphi_0,\pi_0);(f^1,p^1)),\sigma((\varphi_0,\pi_0);(f^2,p^2))]
 =i\sigma((f^1,p^1);(f^2,p^2))I$$
and to define (symplectically smeared) time-$t$ sharp-time fields by demanding
\begin{equation}
\label{quantclass}
\sigma((\varphi_t,\pi_t);(f_t,p_t))=\sigma((\varphi_0,\pi_0);(f_0,p_0))
\end{equation}
where $(f_t,p_t)$ is the classical time-evolute of $(f_0,p_0)$. This
$*$-algebra of sharp-time fields may be identified with the (minimal)
field $*$-algebra of the Section \ref{staralgHad}, the $\hat\phi(F)$ of that section being 
identified with $\sigma((\varphi_0,\pi_0);(f,p))$
where $(f, p)$ are the Cauchy data at $t=0$ of $\Delta F$.  (This
identification is of course many-one since $\hat\phi(F)=0$ whenever $F$
arises as $(\Box_g - m^2)G$ for some test function $G\in
C_0^\infty({\cal M})$.)

Specializing momentarily to the case of the free scalar field $(\Box -
m^2)\phi=0$ ($m \ne 0$) in Minkowski space with a flat $t=0$ Cauchy
surface, the ``symplectically smeared'' two-point function of the usual
ground state (``Minkowski vacuum state''), $\omega_0$, is given, in this
formalism, by
\begin{equation} 
\label{mink}
\omega_0(\sigma((\varphi, \pi); (f^1, p^1))\sigma((\varphi, \pi); (f^2,
p^2)))={1\over 2}(\langle f^1|\mu f^2\rangle + \langle
p^1|\mu^{-1}p^2\rangle +i\sigma((f^1,p^1); (f^2,p^2))
\end{equation}
where the inner products are in the {\it one-particle Hilbert space}
${\cal H} = L^2_{\bf C}({\mathbb R}^3)$ and $\mu=(m^2-\nabla^2)^{1/2}$. The
GNS representation of this state may be concretely realised on the
familiar {\it Fock space} ${\cal F}({\cal H})$ over ${\cal H}$ by
$$\rho_0(\sigma((\varphi, \pi); (f,p)))= -i(\hat a^\dagger(\a)-(\hat
a^\dagger(\a))^*)$$ where $\a$ denotes the element of $\cal H$
$$\a={(\mu^{1/2}f+i\mu^{-1/2}p)\over\sqrt 2}$$
(we note in passing that, if we equip $\cal H$ with the symplectic
form $2\rm{Im}\langle\cdot|\cdot\rangle$, then $K: (f,p)\mapsto \a$ is a
symplectic map -- see Footnote \ref{K})
and $\hat a^\dagger(\a)$ is the usual smeared {\it creation operator}
(=``$\int \hat a^\dagger({\bf x})\a({\bf x})d^3{\bf x}$'') on
${\cal F}({\cal H})$ satisfying
$$[(\hat a^\dagger(\a^1))^*, \hat a^\dagger(\a^2)]=\langle
\a^1|\a^2\rangle_{\cal H} I.$$
The usual (smeared) {\it annihilation operator}, $\hat a(\a)$, is $(\hat
a^\dagger(C\a))^*$ where $C$ is the natural complex conjugation,
$\a\mapsto \a^*$ on $\cal H$.  Both of these operators annihilate the
{\it Fock vacuum vector} $\Omega^{\cal F}$. In this representation, the
one parameter group of time-translation automorphisms
\begin{equation} 
\label{auto}
\alpha(t): \sigma((\varphi_0, \pi_0); (f,p))\mapsto
\sigma((\varphi_t, \pi_t); (f,p))
\end{equation}
is implemented by $\exp(-iHt)$ where $H$ is the second quantization of
$\mu$ (i.e.\ the operator otherwise known as $\int\mu(k)\hat
a^\dagger(k)\hat a(k)d^3k$) on ${\cal F}({\cal H})$.

The most straightforward (albeit physically artificial) situation
involving ``particle creation'' in a curved spacetime concerns a
globally hyperbolic spacetime which, outside of a compact region, is
isometric to Minkowski space with a compact region removed -- i.e.\ to a
globally hyperbolic spacetime which is flat except inside a localized
``bump'' of curvature.  See Figure 1.  (One could also add a potential, $V$, to $m^2$ in (\ref{kg}) provided it is supported inside the bump.)   On the field algebra (defined as in Section \ref{staralgHad}) of such a spacetime, there will be an {\it in} vacuum state (which may be identified with the
Minkowski vacuum to the past of the bump) and an {\it out} vacuum state
(which may be identified with the Minkowski vacuum to the future of the
bump) and one expects e.g.\ the ``in vacuum'' to arise as a many particle
state in the GNS representation of the ``out vacuum'' corresponding to
the creation of particles out of the vacuum by the bump of curvature.

\begin{figure}[h]
\label{fig:bump}
\centering
\includegraphics[trim = 3cm 21cm 2cm 4cm, clip]{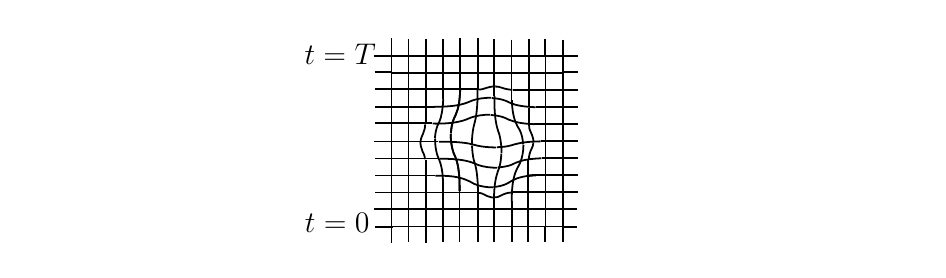}
\caption{\small A spacetime which is flat outside of a compact bump of curvature.}
\end{figure}

\bigskip

In the formalism of this section, if we choose our global time
coordinate on such a spacetime so that, say, the $t=0$ surface is to the
past of the bump and the $t=T$ surface to its future, then the single
automorphism $\alpha(T)$ (defined as in (\ref{auto})) encodes the
overall effect of the bump of curvature on the quantum field and one can
ask whether $\alpha(T)$ is implemented by a unitary operator in the GNS
representation of the Minkowski vacuum state (\ref{mink}).

This question may be answered by referring to the real-linear map ${\cal
T}: {\cal H}\rightarrow {\cal H}$ which sends
$\a_T=2^{-1/2}(\mu^{1/2}f_T+i\mu^{-1/2}p_T)$ to
$\a_0=2^{-1/2}(\mu^{1/2}f_0+i\mu^{-1/2}p_0)$.  By the conservation in
time of $\sigma$ and the symplecticity, noted in passing above, of the
map $K:(f,p)\mapsto \a$, this satisfies the defining relation
$$\rm{Im}\langle {\cal T} a^1|{\cal T} a^2\rangle=\rm{Im}\langle
a^1|a^2\rangle$$
of a {\it classical Bogoliubov transformation}.  Splitting ${\cal T}$
into its complex-linear and complex-antilinear parts by writing $${\cal
T} = \alpha + \beta C$$ where $\alpha$ and $\beta$ are complex linear
operators,  this relation may alternatively be expressed in terms of the
pair of relations $$\alpha^*\alpha-\bar\beta^*\bar\beta=I, \quad
\bar\alpha^*\bar\beta =  \beta^*\alpha$$ where $\bar\alpha=C\alpha C$,
$\bar\beta=C\beta C$.

We remark that there is an easy-to-visualize equivalent way of defining
$\alpha$ and $\beta$ in terms of the analysis, to the past of the bump,
into {\it positive and negative-frequency parts} of {\it complex} solutions to
(\ref{kg}) which are purely {\it positive-frequency} to the future of
the bump.  In fact, if, for any element $\a\in {\cal H}$, we identify
the positive frequency solution to the Minkowski-space Klein Gordon
equation
$$\phi^{\mathrm{pos}}_{\mathrm{out}}(t, {\bf x})=
((2\mu)^{-1/2}\exp(-i\mu t)\a)({\bf x})$$
with a complex solution to (\ref{kg}) to the future of the bump, then
(it may easily be seen) to the past of the bump, this same solution will
be identifiable with the (partly positive-frequency, partly
negative-frequency) Minkowski-space Klein-Gordon solution
$$\phi_{\mathrm{in}}(t, {\bf x})=\left ((2\mu)^{-1/2}\exp(-i\mu t\right )
\alpha\a)({\bf x}) +
\left ((2\mu)^{-1/2}\exp(i\mu t)\bar\beta\a\right )({\bf x})$$
and this could be taken to be the defining equation for the operators $\alpha$
and
$\beta$.

It is known (by a 1962 theorem \cite{Shale} of Shale) that the automorphism
(\ref{auto}) will be unitarily implemented if and only if $\beta$ is a
Hilbert-Schmidt operator on $\cal H$.  Wald \cite{Wald79}, in case $m\ge 0$ and
Dimock \cite{Dimock79}, in case $m\ne 0$ have verified that this condition is
satisfied in the case of our bump-of-curvature situation.  In that case,
if we denote the unitary implementor by $U$, we have the results

\begin{description}

\item{(i)} The expectation value $\langle U\Omega|N(\a)U\Omega\rangle_{{\cal
F}(\cal H)}$ of the number operator, $N(\a)=\hat a^\dagger(\a)\hat a(\a)$,
where $\a$ is a normalized element of ${\cal H}$, is equal to
$\langle\beta\a|\beta\a\rangle_{\cal H}$.

\item{(ii)} First note that there exists an orthonormal basis of
vectors,
$e_i$, ($i=1\dots\infty$), in $\cal H$ such that the (Hilbert
Schmidt) operator $\bar\beta^*\bar\alpha^{*-1}$ has the canonical form
$\sum_i \lambda_i \langle Ce_i|\cdot\rangle |e_i\rangle$.  We then have (up to
an
undertermined phase)
$$U\Omega=N\exp\left (-{1\over
2}\sum_i\lambda_i\hat a^\dagger(e_i)\hat a^\dagger(e_i)\right) \Omega.$$
where the normalization constant, $N$, is chosen so that $\|U\Omega\|=1$.
This formula makes manifest that the particles are created in pairs.

\end{description}

\noindent
We remark that, identifying elements, $\a$, of $\cal H$ with
positive-frequency solutions (below, we shall call them ``modes'') as
explained above, Result (i) may alternatively be expressed by saying
that the expectation value, $\omega_{\mathrm{in}}(N(\a))$,  in the {\it
in-vacuum state} of the  occupation number, $N(\a)$, of a {\it
normalized mode}, $\a$, to the future of the bump is given by
$\langle\beta\a|\beta\a\rangle_{\cal H}$.

The formalism and results (i) and (ii) above generalize from spacetimes which are flat outside of a compact bump of curvature to spacetimes which are only asymptotically flat in suitable senses.  In fact much work has been done on classical scattering theory for linear field equations such as (\ref{kg}) and on the implications for the quantum theory.  At a heuristic level, one will still have notions of complex classical solutions which are positive frequency asymptotically towards the past/future, and, in consequence, one will have well-defined asymptotic notions of in and out vacua and of in and out particles built on those vacua.    At a rigorous algebraic level, say for Equation (\ref{kg}), once one has established (See e.g.\ \cite{Kay82} and \cite{DimockKaySchwI}) the existence of classical wave (or `M\o ller') operators $\Omega^\pm$ which can be thought of as mapping solutions, $\phi_0$, of a \emph{comparison Minkowski-space Klein-Gordon equation} to solutions 
$\Omega^\pm\phi_0$ in the spacetime of interest\footnote{Here we assume we can identify the manifold of Minkowski space with the manifold of the spacetime of interest.   When the latter has a different topology as e.g.\ in the case of exterior Schwarzschild, this is of course not possible globally but one may overcome this with suitable mollified partial identifications -- see e.g.\  around \cite[Equation (4.1)]{DimockKaySchwI}.   \cite{DimockKaySchwI} also explains that there is considerable leeway in the way these (partial) identifications are made.  Note also that, for the massive Klein-Gordon field, the interaction of the Schwarzschild background with the Klein-Gordon field is closely analogous to the interaction of a long-range external Coulomb potential with a charged field in Minkowski space and therefore the wave operators constructed in \cite{DimockKaySchwI} are suitably Dollard-modified [see \cite{DucDyb}] wave operators.} which look like $\phi_0$ at late/early times (and once one has also established asymptotic completeness) then one may, e.g., define {\it out} and {\it in} fields according to $\tilde W_\mathrm{out/in}(\phi_0) = \tilde W(\Omega^\pm\phi_0)$ and {\it out} and {\it in} vacuum states $\omega_\mathrm{out/in}$ according to $\omega_{out/in}(\tilde W(\Omega^\pm\phi_0)) = \omega_0(\tilde W_0(\phi_0))$ where $\tilde W_0$ is the Weyl operator and $\omega_0$ the usual vacuum state for the Minkowski space Klein-Gordon equation.  We shall say more about classical and quantum scattering theory on spacetimes related to black holes in Section \ref{blackholes}.

Also, in, e.g.\ cosmological, models where
the background spacetime is slowly-varying in time, one can define
approximate {\it adiabatic} notions of classical positive frequency
solutions, and hence also of quantum ``vacuum'' and ``particles'' at
each finite value of the cosmological time.  But, at times where the
gravitational field is rapidly varying, one does not expect there to be
any sensible notion of ``particles''.  And, in a rapidly
time-varying background gravitational field which never settles down
one does not expect there to be any sensible particle interpretation of
the theory at all. To understand these statements, it suffices to consider the
$1+0$-dimensional Klein-Gordon equation with an external potential $V$:
$$\left (-{d^2\over dt^2}-m^2-V(t)\right )\phi=0$$ 
which is of course a system of one degree of freedom, mathematically
equivalent to the harmonic oscillator with a time-varying angular
frequency $\varpi(t)=(m^2+V(t))^{1/2}$. One could of course express its
quantum theory in terms of a time-evolving Schr\"odinger wave function
$\Psi(\varphi, t)$ and attempt to give this a particle interpretation at
each time, $s$, by expanding $\Psi(\varphi, s)$ in terms of the harmonic
oscillator wave functions for a harmonic oscillator with some particular
choice of angular frequency.  But the problem is, as is easy to convince
oneself, that there is no such good choice.   For example, one might
think that a good choice would be to take, at time $s$, the set of
harmonic oscillator wave functions with angular frequency $\varpi(s)$.
(This is sometimes known as the method of ``instantaneous
diagonalization of the Hamiltonian'').   But suppose we were to apply
this prescription to the case of a smooth $V(\cdot)$ which is constant
in time until time $0$ and assume the initial state is the usual vacuum
state. Then at some positive time $s$, the number of particles predicted
to be present is the same as the number of particles predicted to be
present on the same prescription at all times after $s$ for a $\hat
V(\cdot)$ which is equal to  $V(\cdot)$ up to time $s$ and then takes
the constant value $V(s)$ for all later times (see Figure 2).  But $\hat V(\cdot)$
will generically have a sharp corner in its graph (i.e.\ a discontinuity in
its time-derivative) at time $s$ and one would expect a large part of
the particle production in the latter situation to be accounted for by
the presence of this sharp corner -- and therefore a large part of the
predicted particle-production in the case of $V(\cdot)$ to be spurious.

\begin{figure}[h]
\label{fig:Vgraph}
\centering
\includegraphics[scale=0.5,trim = 7cm 16cm 0cm 4cm, clip]{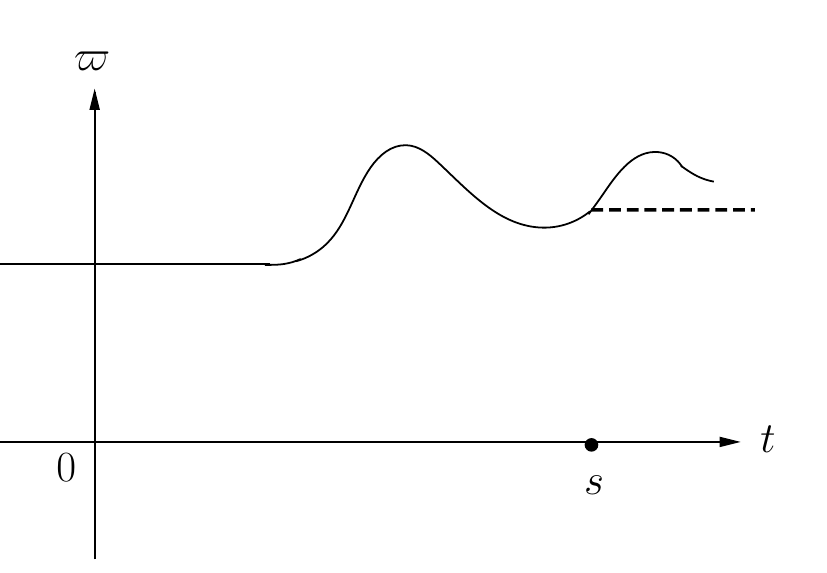}
\caption{\small Plots of $\varpi$ against $t$ for the two potentials
$V$ (continuous line) and $\hat V$ (continuous line up to $s$ and then dashed
line) which play a r\^ole in our critique of ``instantaneous diagonalization''.}
\end{figure}

\bigskip

Back in $1+3$ dimensions, even where a good notion of particles is
possible, it depends on the choice of time-evolution, as is dramatically
illustrated by the Unruh effect (see Section \ref{blackholes}).

\section{\label{stress} Theory of the Stress-Energy Tensor}

To orient ideas, consider first the  free (minimally coupled) scalar
field,  $(\Box - m^2)\phi=0$, in Minkowski space.  If one quantizes this
system in the usual Minkowski-vacuum representation, then the
expectation value of the {\it renormalized stress-energy tensor}  (which
in this case is the same thing as the {\it normal ordered stress-energy
tensor}) in a vector state $\Psi$ in the Fock space will be given by the
formal {\it point-splitting} expression
$$\langle \Psi|T_{ab}(x)\Psi\rangle=\lim_{(x_1,x_2)\rightarrow (x,x)}
\left (\partial_a^1\partial_b^2-{1\over 2}\eta_{ab}(\eta^{cd}\partial_c^1
\partial_d^2+m^2)\right)$$
\begin{equation} 
\label{renvec}
\left(\langle\Psi|\rho_0(\phi(x_1)\phi(x_2))\Psi\rangle
-\langle\Omega^{\cal F}|\rho_0(\phi(x_1)\phi(x_2))\Omega^{\cal
F}\rangle\right)
\end{equation}
where $\eta_{ab}$ is the usual Minkowski metric and $\Omega^{\cal
F}$ the Fock space vacuum vector.  A sufficient condition
for the limit here to be finite and well-defined would, e.g., be for
$\Psi$ to consist of a (normalised) finite superposition of $n$-particle
vectors of form $\hat a^\dagger(\a_1),\dots ,\hat
a^\dagger(\a_n)\Omega^{\cal F}$ where the smearing functions $\a_1,\dots
,\a_n$ are all $C^\infty$ elements of $\cal H$ (i.e.\ of $L^2_{\bf C}({\bf
\mathbb R}^3)$).  The reason this works is that the two-point function in such
states shares the same short-distance singularity as the
Minkowski-vacuum two-point function.  For exactly the same reason, one
obtains a well-defined finite limit if one defines the expectation value
of the stress-energy tensor in any physically admissible quasi-free
state by the expression
\begin{equation} 
\label{ren}
\omega(T_{ab})(x))=\lim_{(x_1,x_2)\rightarrow (x,x)}
\left (\partial_a^1\partial_b^2-{1\over 2}\eta_{ab}(\eta^{cd}\partial_c^1
\partial_d^2+m^2)\right)\left(\omega(\phi(x_1)\phi(x_2))
-\omega_0(\phi(x_1)\phi(x_2))\right ).
\end{equation}
This latter point-splitting formula generalizes to a definition for the
{\it expectation value of the renormalized stress-energy tensor} for an
arbitrary physically admissible quasi-free state (or indeed for an
arbitrary state whose two point function has {\it Hadamard form} -- i.e.\
whose anticommutator function satisfies Condition (d) of Section 2) on
the minimal field algebra and to other linear field theories (including
the stress tensor for a conformally coupled linear scalar field) on a
general globally hyperbolic spacetime (and the result obtained agrees
with that obtained by other methods, including {\it dimensional
regularization} and {\it zeta-function regularization}).  However, the
generalization to a curved spacetime involves a number of important new
features which we next briefly list.  (See (Wald, 1978) for the original way in which this was done and see D\'ecanini and Folacci \cite{DecFol} for a technically slightly different and, in some ways simpler, alternative\footnote{In (Wald, 1978), the locally constructed Hadamard two-point function which is used as the subtraction piece, say $H(x,x')$, is defined by demanding that the first term in a Hadamard expansion $\sum_{n=0}^\infty w_n\sigma^n$ for the $w$ term in the Hadamard formula (see item (d) in Section \ref{staralgHad} but note that Wald's $w$ corresponds to what in (d) and in \cite{DecFol} would be $Uw$) takes the same value as in Minkowski space.   This results in a subtraction term which, as explained in (Wald, 1978) satisfies (\ref{kg}) in one of its arguments, but (generically) not the other, and which fails to be symmetric.   On the other hand, in \cite{DecFol}, D\'ecanini and Folacci (who work with the expectation value in some state of the time-ordered product of $\phi(x)$ with $\phi(x')$ rather than the expectation value of the anticommutator as in (Wald, 1978) and the present paper, but this makes little difference -- other than a factor of 2 -- for the stress-energy tensor) take the subtraction piece, say $H'(x,x')$, to be simply the first two terms in the expression in (d) (for some choice of length scale).   This is symmetric but fails (generically) to satisfy (\ref{kg}) in {\it either} of its arguments.  Nevertheless the procedures in (Wald, 1978) and \cite{DecFol} both give the same result for the stress-energy tensor (up to the usual ambiguity in local geometrical terms) and the same formula for the trace anomaly.   See also \cite{MorettiCommStress} for yet another alternative approach which also spells out the generalization to other dimenions.  Note that D\'ecanini and Folacci \cite{DecFol} also discuss the Hadamard form in dimensions other than 4.  Also see \cite{BalaWin} for the generalization -- in the case of a charged scalar field -- of what is done in \cite{DecFol} to include an external electromagnetic field.}.)

First, the subtraction term which replaces
$\omega_0(\phi(x_1)\phi(x_2))$ is, in general,  not the expectation
value of $\phi(x_1)\phi(x_2)$ in any particular state, but rather a
particular {\it locally constructed Hadamard two-point function} whose
physical interpretation is more subtle; the renormalization is thus in
general not to be regarded as a normal ordering.  Second, the immediate
result of the resulting limiting process will not be covariantly
conserved and, in order to obtain a covariantly conserved quantity, one
needs to add a particular {\it local geometrical correction term}.  The
upshot of this is that the resulting expected stress-energy tensor is
covariantly conserved but possesses a (state-independent) {\it anomalous
trace}.  In particular, for a massless conformally coupled linear scalar
field, one has (for all physically admissible quasi-free states,
$\omega$) the {\it trace anomaly formula}
$$\omega(T^a_a(x))=(2880\pi^2)^{-1}(C_{abcd}C^{abcd}+R_{ab}R^{ab}-{1\over
 3}R^2)$$
 -- plus an arbitrary multiple of $\Box R$.  In fact, in general, the
thus-defined renormalized stress-energy tensor operator (see below) is
only defined up to a {\it finite renormalization ambiguity} which
consists of the addition of arbitrary multiples of the functional
derivatives with respect to $g_{ab}$ of the four quantities
$$I_n=\int_{\cal M} F_n(x)|\det(g)|^{1/2}d^4x,$$
$n=1\dots 4$, where $F_1=1$, $F_2=R$, $F_3=R^2$, $F_4=R_{ab}R^{ab}$.   In the
Minkowski-space case, only the first of these ambiguities arises and it
is implicitly resolved in the formulae (\ref{renvec}), (\ref{ren})
inasmuch as these effectively incorporate the {\it renormalization
condition} that $\omega_0(T_{ab})=0$.  (For the same reason, the
locally-flat example we give below has no ambiguity.)

One expects, in both flat and curved cases, that, for test functions,
$F\in C_0^\infty({\cal M})$, there will exist operators $T_{ab}(F)$
which are {\it affiliated to} the net of local  $W^*$-algebras referred
to in Section 2 and that it is meaningful to write 
$$\int_{\cal
M}\omega(T_{ab}(x))F(x)|\det(g)|^{1/2}d^4x=\omega(T_{ab}(F))$$  
provided that, by $\omega$ on the right-hand side, we understand the
extension of $\omega$ from the Weyl algebra to this net.  ($T_{ab}(F)$
is however not expected to belong to the minimal algebra or be
affiliated to the Weyl algebra.)

An interesting simple example of a renormalized stress-energy tensor
calculation is the so-called {\it Casimir effect} calculation for a
linear scalar field on a (for further simplicity, $1+1$-dimensional)
{\it timelike cylinder spacetime} of radius $R$ (see Figure 3). This
spacetime is globally hyperbolic and stationary and, while locally flat,
globally distinct from Minkowski space.  As a result, while  -- provided
the regions ${\cal O}$ are sufficiently small (such as the diamond
region in Figure 3) -- elements ${\cal A}({\cal O})$ of the minimal net
of local algebras on this spacetime will be identifiable, in an obvious
way, with elements of the minimal net of local algebras on Minkowski
space, the stationary ground state $\omega_{\mathrm{cylinder}}$ will, when
restricted to such thus-identified regions, be distinct from the
Minkowski vacuum state $\omega_0$.  The resulting renormalized
stress-energy tensor (as first pointed out in (Kay, 1979), defineable,
once the above identification has been made, exactly as in (\ref{ren}))
turns out to be non-zero and, interestingly, to have an (in the natural
coordinates, constant) {\it negative energy density} $T_{00}$.  In fact:
$$\omega_{\mathrm{cylinder}}(T_{ab})={1\over 24\pi R^2}\eta_{ab}.$$

\begin{figure}[h]
\label{fig:casimir}
\centering
\includegraphics[scale=1.4, trim = 6cm 19.5cm 10cm 7cm, clip]{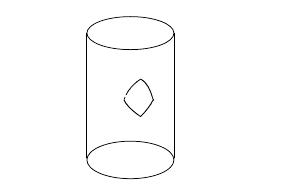}
\caption{\small The time-like cylinder spacetime of radius $R$ with a
diamond region isometric to a piece of Minkowski space.  See (Kay, 1979).}
\end{figure}

\section{\label{AQFT} \textit{More about the Intersection of QFTCST with AQFT and the 
Fewster-Verch No-Go Theorem}}

We have already indicated in the introduction that there is a close interrelationship between QFTCST and algebraic quantum field theory (AQFT).   In recent years, this connection has got even closer.  See e.g.\ the collection (Brunetti et al., 2015).   This process was greatly accelerated by two conceptual revolutions, the first of which centred on the introduction in 1992 by Radzikowski (see Section \ref{staralgHad}) of the microlocal spectrum condition  -- which was susceptible \cite{Verch1999,BrunFred2000} to generalization to interacting fields. The second conceptual revolution was the adoption of a new perspective in which, rather than focus on quantum field theory on any single fixed curved spacetime, one asks instead about the simultaneous specification of a quantum field theory on a whole family of (at least at first, globally hyperbolic) spacetimes.  More precisely one defines a \emph{locally covariant quantum field theory} to consist of a covariant functor between \emph{the category of globally hyperbolic spacetimes with isometric embeddings} as morphisms and \emph{the category of $*$-algebras with unital injective $*$-monomorphisms} as morphisms.   \footnote{\label{helical} The morphisms on the spacetime side need to be restricted i.a.\ so that if one spacetime is embedded as a subspacetime of another another, then their causal structures are compatible.  The need for this is illustrated by the helical strip example of \cite{KayFlocality}.}   A key paper which might be said to be a manifesto for this approach is (Brunetti et al., 2003) (see also the review article \cite{FewsterVerch2015}) although some of the ingredients in the approach had been apparent in earlier works.  In particular, Dimock (Dimock, 1980) had advocated a related shift of perspective while the papers (Kay, 1979) and \cite{KayFlocality} (see Footnote \ref{helical}) had anticipated some aspects of the r\^ole that the embedding of one spacetime inside another might play.   Also some results had already been obtained in the locally covariant framework before (Brunetti et al., 2003) appeared.   See e.g.\ Verch's paper  \cite{VerchSpinStat} on generalizing the spin statistics theorem to curved spacetimes. 

Amongst the several interesting new research directions that flowed from the adoption of the locally covariant approach to QFTCST, let us mention the work of Fewster and Verch \cite{FewsterVerch2012a} which shows that a locally covariant QFT will be ``the same in all spacetimes''\footnote{This is what Fewster and Verch call their `same physics in all spacetimes' or `SPAS' property.} in a certain natural sense provided it satisfies a condition that they introduce and call  \emph{dynamical locality}.   They show that this condition is satisfied by a wide class of quantum field theories, although they also point out that there are some quantum field theories for which it doesn't hold -- with interesting consequences.   For those theories for which dynamical locality holds (together with some other mild technical conditions), which includes \cite{FewsterVerch2012b} the quantum theory of the massive minimally coupled Klein Gordon equation (\ref{kg}), it is proved in \cite{FewsterVerch2012a} (see also \cite{FewArtState}) i.a.\ that \emph{it is impossible to make a locally covariant choice of preferred state in all spacetimes} (in the above category of spacetimes).   We shall call this the \textit{Fewster-Verch no-go theorem}.

To gain some insight into why the Fewster-Verch no-go theorem holds in the case of (for simplicity, the 1+1-dimensional version of) Equation (\ref{kg}) let us assume that, if there were such a covariant choice of preferred state, then, in the case of (1+1-dimensional) Minkowski space, it would coincide with the usual (Poincar\'e invariant) ground state, $\omega_0$, and that, in the case of the (1+1-dimensional) timelike cylinder spacetime (see Section \ref{stress}) it would coincide with the stationary ground state $\omega_{\mathrm{cylinder}}$.  (See Section \ref{stress}.) Then, if there were a covariant choice of preferred state on all spacetimes, the preferred state on a diamond region of Minkowski space (see Figure 3), viewed as a globally hyperbolic spacetime in its own right, would have, at the same time, to coincide with the restriction to it of the Minkowski vacuum when we isometrically embed the diamond in Minkowski space, and with the restriction to it of $\omega_{\mathrm{cylinder}}$ when we isometrically embed the diamond in the timelike cylinder in the way illustrated in Figure 3.  But, as we saw in Section \ref{stress}, these states cannot coincide since, e.g., they have different expectation values for the renormalized stress energy tensor at any point in the diamond.  A contradiction!

\section{\label{blackholes} Hawking and Unruh effects}

The original calculation by Hawking (Hawking, 1975) concerned a model spacetime
for a star which collapses to a black hole. For simplicity, we shall
only discuss the spherically-symmetric case.  (See Figure 4.) 

\begin{figure}[h]
\label{fig:collapse}
\centering
\includegraphics[trim = 6cm 21cm 7cm 5cm, clip]{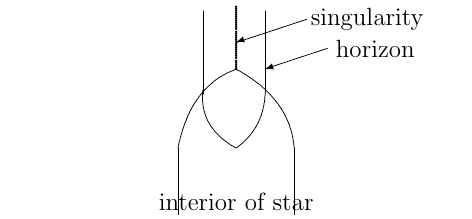}
\caption{\small The spacetime of a star collapsing to a spherical black
hole.}
\end{figure}

\bigskip

Adopting a similar ``mode'' viewpoint to that mentioned after Results
(i) and (ii) in Section \ref{particle}, the result of the calculation may be stated
as follows: For a real linear scalar field satisfying (\ref{kg}) with
$m=0$ (and $V=0$) on this spacetime, the expectation value
$\omega_{\mathrm{in}}(N(\a_{\varpi,\ell}))$ of the occupation number of a
one-particle outgoing mode $\a_{\varpi, \ell}$ localized (as far as a
normalized mode can be) around $\varpi$ in angular-frequency-space and
about retarded time $v$ and with angular momentum ``quantum number''
$\ell$, in the in-vacuum state (i.e.\ on the minimal algebra for a real
scalar field on this model spacetime)  $\omega_{\mathrm{in}}$ is, at late
retarded times, given by the formula
$$\omega_{\mathrm{in}}(N(\a_{\varpi,\ell}))={\Gamma(\varpi,\ell)\over
\exp(8\pi M\varpi)-1}$$
where $M$ is the mass of the black hole and the {\it absorption factor}
(alternatively known as {\it grey body factor})  $\Gamma(\varpi, \ell)$
is equal to the norm-squared of that part of the one-particle mode,
$\a_{\varpi,\ell}$, which, viewed as a complex positive frequency
classical solution propagating backwards in time from late retarded
times, would be absorbed by the black hole.  This calculation can be
understood as an application of Result (i) of Section \ref{particle} (even though the
spacetime is more complicated than one with a localized ``bump of
curvature'' and even though the relevant overall time-evolution will not
be unitarily implemented, the result still applies when suitably
interpreted) and the heart of the calculation is an asymptotic estimate
of the relevant ``$\beta$'' Bogoliubov coefficient which turns out to be
dependent on the geometrical optics of rays which pass through the star
just before the formation of the horizon.   This result suggests that
the in-vacuum state is indistinguishable at late retarded times from a
state of black-body radiation at the {\it Hawking temperature},
$T_{\mathrm{Hawking}}=1/8\pi M$, in Minkowski space from a black-body with
the same absorption factor. 

The above picture of Hawking radiation for the spacetime of stellar collapse to a black hole  was confirmed by further work by many
authors. Much of that work, as well as the original result of Hawking,
was partially heuristic but considerable effort has been made over the nearly 50 years since Hawking's work to obtain rigorous results.

A first such rigorous result is that  \cite{FredHaag} of Fredenhagen and Haag who showed for the (for simplicity, massless) Klein Gordon equation (\ref{kg}) and for a model  of a stationary detector far from a spherical star collapsing to a black hole, and assuming that the quantum field reflects off the surface of the collapsing star, that the ``counting rate'' of the detector will be consistent with Hawking's prediction as a consequence of the requirement that the two point function of the state will have the right scaling limit (a weaker condition than the Hadamard condition) at the sphere where the surface of the collapsing star crosses the Schwarzschild radius.

It is also, however, desirable to show, in a scattering theory sense, that an {\it in} vacuum state (say viewed as a state on a comparison copy of Minkowski space, or, in the case of (conformally coupled) massless fields at ${\cal I}^-$\footnote{Here ${\cal I}^+$ and ${\cal I}^-$ -- pronounced {\it scri plus} and {\it scri minus} -- refer to future and past null infinity in the sense of Penrose's conformal compactification of asymptotically flat spacetimes (see e.g.\ \cite{HawkingEllis}).   In the case of massive fields, the notions of ${\cal I}^+$ and ${\cal I}^-$ are of little use since, when multiplied by the appropriate conformal factors, classical solutions won't have finite restrictions to them, but one may instead consider the inverse action of (Dollard-modified) {\it in} and {\it out} wave operators which map full solutions to solutions of a comparison Klein-Gordon equation on Minkowski space.   See around the brief mention of the relation between {\it in} and {\it out} fields and classical wave operators in Section \ref{particle}.}) evolves into an {\it out} state (say viewed as a state in a comparison copy of Minkowski space, or, in the case of massless fields, at ${\cal I}^+$) which is identical with the state of thermal radiation predicted by Hawking.

Let us remark here that one expects that the assumption of an {\it in} vacuum state in the Hawking effect is actually not so important and that, e.g.\ a multiparticle state built on the {\it in} vacuum will lead to similar behaviour at late times.  This is because one expects the Hawking radiation from such an {\it in} state will differ from the Hawking radiation from the {\it in} vacuum by transients which disappear completely at ``future infinity''.    This point was already made by Wald in \cite{WaldTrans} and is also implicit in the Fredenhagen-Haag result mentioned above since one would expect  multiparticle {\it in} states (with suitable smoothness properties) to have the same scaling limit for their two-point functions as the {\it in} vacuum.  Another point (also made in \cite{FredHaag} and, earlier, by Leahy and Unruh in \cite{LeahyUnruh}) is that for an interacting quantum field theory there is no reason to expect the Hawking radiation due to stellar collapse to be exactly thermal (i.e.\ exactly a Gibbs state on modes which emanate from the black hole).   This is to be contrasted with the Hartle-Hawking-Israel state (see below) on exterior Schwarzschild whose Gibbs-state nature is expected to generalize to interacting fields.

For the case of a massive or massless Klein Gordon field and spherical collapse, a first step was taken towards a scattering theoretic understanding of the Hawking effect by Dimock and Kay \cite{DimockKaySchwI} who, after developing a suitable classical scattering theory on exterior Schwarzschild, used that, i.a.,  to give a rigorous definition of the \emph{Unruh state}.  (In the massive case there was an assumption of asymptotic completeness which has, however, since been proven \cite{BachAsymp} by Bachelot.  In the massless case asymptotic completeness was proven by Dimock in \cite{DimAsymp}.)   In the case of massless fields, this is an important state on exterior Schwarzschild introduced in (Unruh, 1976) which is an {\it in} vacuum at the past horizon (with respect to the affine parameter of its null generators\footnote{which is the same thing as a thermal state at the Hawking temperature with respect to Lorentz boosts -- see the discussion below about the \emph{Unruh effect} and see (Kay and Wald, 1991).}) as well as at ${\cal I}^-$ and which Unruh argued should coincide with the state of Hawking radiation on the exterior of a collapsing star on ${\cal I}^+$ (when the ${\cal I}^+$ of the spacetime of the collapsing star is identified with the ${\cal I}^+$ of exterior Schwarzschild).    Dimock and Kay proved in both massive and massless cases that their rigorously defined Unruh state at late times, viewed as a state for the quantized Equation (\ref{kg}) on a comparison copy of Minkowski space\footnote{In the scattering theory of \cite{DimockKaySchwI}, there are actually two comparison systems.  There is the ordinary, 4-dimensional Klein-Gordon equation on 4-dimensional Minkowski space to capture the behaviour at large distances and early or late times and there is the 2-dimensional massless wave equation on a 2-dimensional Minkowski space  producted with a 2-sphere to capture early and late-time behaviour near the Schwarzschild radius.  In the main text here, we are referring to the former.} is indeed the state of Hawking radiation that one would expect -- namely (see \cite{DimockKaySchwI} for details) the state of black body radiation from a black body with an `absorption operator' defined appropriately in terms of the relevant classical wave operators.\footnote{In showing this, \cite{DimockKaySchwI} essentially give an appropriate rigorous notion of the `grey body' factors mentioned earlier.}  However, as was written at the end of the introduction of \cite{DimockKaySchwI}, ``a complete treatment of the Hawking effect would [also] require a parallel discussion of the scattering problem for a generic collapsing star and a discussion of the precise sense in which the Unruh state approximates the {\it in} vacuum state for the spacetime of a collapsing star.''

Complete treatments of the Hawking effect have since been given by proving directly on the spacetimes of models for a collapsing star (i.e.\ without consideration of the sense in which the Unruh state approximates the {\it in} vacuum state for the spacetime of a collapsing star) that the {\it in} vacuum state coincides with the expected state of Hawking radiation either on ${\cal I}^+$ (for massless fields) or on the {\it out} auxiliary copy of Minkowski space (for massive or massless fields).   The first such result \cite{BachHawk} was obtained by Bachelot for the massive or massless Klein Gordon field on a spacetime describing a spherical collapse which is stationary in the past in which the field does not penetrate the surface of the star but rather reflects off it.   A similar result \cite{Melnyk} for the charged massive Dirac field was obtained by Melnyk.

More recently, Alford \cite{AlfordPhD} (see also \cite{AlfordPaper1}) has given a treatment of the Hawking effect for the massless Klein-Gordon equation on the spacetime of the Oppenheimer-Snyder solution, treating the case where the quantum field penetrates the region occupied by the collapsing dust ball as well as the case where it reflects off its surface.  Even more recently, in \cite{AlfordPaper2,AlfordPaper3}, Alford has extended the treatment (still of the chargeless massless Klein-Gordon equation) in the reflective case to more general collapsing spherical dust clouds which may themselves be electrically charged and which collapse to both subextremal and extremal Reissner Nordstr\"om black holes.   Amongst other results, he proves a mathematical result that confirms that, as would be expected, in the extremal case, the Hawking radiation is transient.

Finally, let us mention that there has  been a lot more work on the scattering (and related properties) of classical linear waves on black hole backgrounds which, while it does not explicitly explore quantum corollaries that may be derived from the classical results (see Footnote \ref{QuantCoroll}) is no doubt susceptible to interesting such corollaries.   See for example \cite{DafRodLect,DafRodShl,HafNicScatt,Batic}.

The Hawking black-hole radiation result suggests that there is something fundamentally ``thermal'' about quantum fields on black-hole backgrounds and our understanding of this is deepened by consideration of quantum fields, not on the spacetime of stellar collapse, but on the maximally extended Schwarzschild spacetime (a.k.a.\  the {\it Kruskal} or  {\it Kruskal-Szekeres} spacetime see Figure 5).  In particular, the theorems in the two papers (Kay and Wald, 1991) and (Kay, 1993), combined together, tell
us that there is a unique state on the Weyl algebra for Kruskal which is invariant under the
{\it Schwarzschild isometry group} and whose two-point function has
Hadamard form.   
(Let us also mention here that a gap in the proofs in (Kay and Wald, 1991) was identified and corrected in Appendix B of \cite{KayLupo}.)  
Moreover, they tell us that this state, when restricted
to a single wedge (i.e.\ the exterior Schwarzschild spacetime) is
necessarily a KMS state at the Hawking temperature.

\begin{figure}[h]
\label{fig:kruskal}
\centering
\includegraphics[scale = 0.5, trim = 3cm 12cm 0cm 4cm, clip]{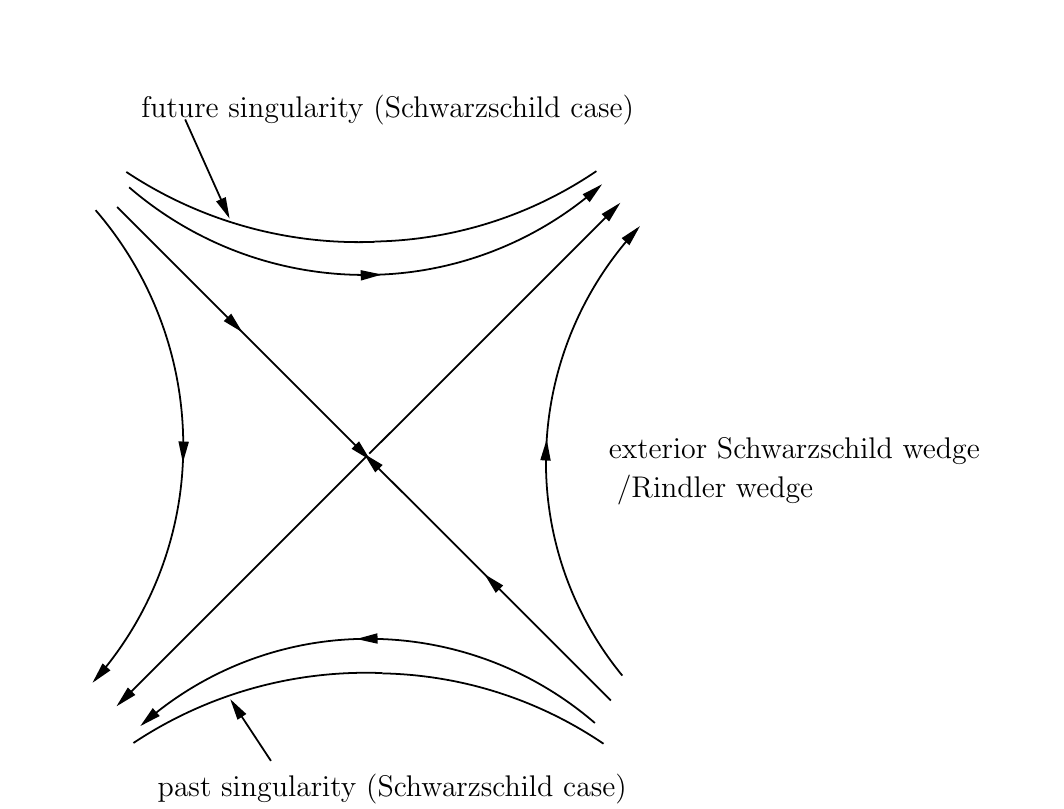}
\caption{\small The geometry of maximally extended Schwarzschild
(/or Minkowski) spacetime.  In the Schwarzschild case, every point
represents a two-sphere (/in the Minkowski case, a two-plane).  The
curves with arrows on them indicate the Schwarzschild time-evolution
(/one-parameter family of Lorentz boosts).  These curves include the (straight
lines at right angles) event horizons (/Killing horizons).}
\end{figure}

\bigskip

This unique state is known as the {\it Hartle-Hawking-Israel state} (\emph{HHI} state).  
These results in fact apply more generally to a wide class of globally
hyperbolic spacetimes with {\it bifurcate Killing horizons} including de
Sitter space -- where the unique state is sometimes called the {\it
Euclidean}  and sometimes the {\it Bunch-Davies} vacuum state (see e.g.\ the recent paper \cite{HiguchiYamamoto2018}) -- as well
as to Minkowski space, in which case the unique state is the usual
Minkowski vacuum state,  the analogue of the exterior Schwarzschild
wedge is a so-called {\it Rindler wedge}, and the relevant isometry
group is a one-parameter family of wedge-preserving Lorentz boosts. In
the latter situation, the fact that the Minkowski vacuum state is a KMS
state (at ``temperature'' $1/2\pi$) when restricted to a Rindler wedge
and regarded with respect to the time-evolution consisting of the
wedge-preserving one-parameter family of Lorentz boosts is known as the
{\it Unruh effect} (Unruh, 1976). See \cite{CrispinoEtAl} for a review.
This latter property of the Minkowski vacuum
in fact generalizes to general {\it Wightman quantum field theories} and
is in fact, as first pointed out by Sewell \cite{Sewell80,Sewell82}, an immediate 
consequence of a combination of the {\it Reeh Schlieder Theorem} (applied to a Rindler wedge) and the {\it Bisognano Wichmann Theorem}.  The latter theorem says that the defining relation (\ref{kms}) of a KMS state holds if, in (\ref{kms}), we
identify the operator $J$ with the complex conjugation which implements
wedge reflection and $H$ with the self-adjoint generator of the unitary
implementor of Lorentz boosts. We remark that the Unruh effect
illustrates how the concept of ``vacuum'' (when meaningful at all) is
dependent on the choice of time-evolution under consideration.  Thus the
usual Minkowski vacuum is a ground state with respect to the usual
Minkowski time-evolution but not (when restricted to a Rindler wedge)
with respect to a one-parameter family of Lorentz boosts; with respect
to these, it is, instead, a KMS state. 

Let us also mention that illuminating heuristic arguments for the existence of the Hartle-Hawking-Israel state, even in the case of interacting field theories, based on formal Euclidean path integrals and Wick rotation were given by Jacobson in \cite{Jacobson94}.

A first rigorous construction of the Hartle-Hawking-Israel state on the Kruskal spacetime for the Klein-Gordon model (\ref{kg}) and the proof of some of its properties was first carried out in \cite{KayDoubleWedge}.  However that construction was not for the full Kruskal spacetime, but rather only for the double exterior Schwarzschild wedge.   In terms of symplectically smeared sharp-time fields on a Cauchy surface through the bifurcation 2-sphere, the smearing functions were bounded away from the bifurcation 2-sphere.    The removal of this restriction and thereby the full construction on the entire Kruskal spacetime, and, more generally, on other spacetimes with \emph{static} bifurcate Killing horizons, was first carried out by Sanders \cite{SandersHHI} based on Euclidean methods and Wick rotation.   Sanders also showed that the thus constructed HHI states are Hadamard.    Also using Euclidean methods and Wick rotation but with a technically different approach based on Calder\'on projectors,  G\'erard \cite{GerardHHI} has more recently obtained similar results where staticity is relaxed to stationarity.  However, the results in \cite{GerardHHI} require the bifurcation surface to be compact.  

As pointed out in \cite{GerardHHI}, the latter result excludes the Kerr spacetime since the exterior wedge of that is of course not globally stationary.  In fact, it was argued in (Kay and Wald, 1991) that because of superradiance -- or relatedly, because the Killing vector which is null on the horizon is spacelike well away from the horizon (see also \cite{PinSanVer}) for the Klein-Gordon equation, (\ref{kg}),
 \emph{there does not exist any isometry-invariant Hadamard state on (the appropriate globally hyperbolic part of) Kerr} and there is also a similar no-go result in (Kay and Wald, 1991) for spacetimes with bifurcate Killing horizons such as Schwarzschild-deSitter which include triple wedges.   Another no-go result of this type was argued for by Kay and Lupo \cite{KayLupo}, namely that there is no stationary Hadamard state if one removes from the Kruskal spacetime the part of the right wedge (see Figure 5) to the right of a surface of constant Schwarzschild radius and imposes, say, Dirichlet boundary conditions on that surface -- unless one removes a similar surface and imposes similar boundary conditions on a `mirror surface' in the left wedge.   This raises interesing questions about what could be the right idealization of a `black hole in a (spherical) box' which plays an important r\^ole in black hole thermodynamics \cite{HawkingBHThermo}.

\section{\label{MoreBH} \textit{More about (Classical and) Quantum Fields on Black Hole Backgrounds}}

A construction has been given \cite{CasalsEtAl} for a Hartle-Hawking-Israel-like state (as well as several other states including an Unruh-like state) for the Dirac equation in a Kerr background and several of its properties obtained.  But while some indications were obtained, it seems to still be unknown whether or not that is everywhere a Hadamard state.   (We remark here in relation to this that the no-go result in (Kay and Wald, 1991) that there is no isometry-invariant Hadamard state for the Klein-Gordon equation on Kerr may well not  generalize to Dirac fields since there is no superradiance for Dirac fields.)

Concerning the rigorous construction of the Unruh state, as far as I am aware, nothing further has been done for the massive Klein-Gordon equation on Schwarzschild since the work of Dimock and Kay -- as supplemented by the asymptotic completeness result of Bachelot -- as we mentioned earlier. In particular, while one expects it to be Hadamard, this seems not to have been established.  (The situation is different, however, for Schwarzschild-de Sitter -- and its electrically charged and rotating generalizations -- see below.)    However, for the massless case, a construction has been given \cite{DappEtAlUnruh} which extends to the full Kruskal spacetime and the resulting Unruh state is shown there to be Hadamard.  See also \cite{OriOttEtAl} for a computation of the two-point function in this state, especially for pairs of points inside the horizon.  A further work \cite{GerHafWro} obtains a similar result for the massless Dirac equation on slowly rotating Kerr based on the classical scattering theory of \cite{HafNicScatt}.

A recent development of great interest concerns the behaviour of quantum fields around the Cauchy horizons of spacetimes such as Reissner Nordstr\"om-de Sitter and Kerr-de Sitter.  For ordinary Reissner-Nordstr\"om and Kerr, as has been much discussed since Penrose first pointed it out in \cite{PenBatelle} in 1968 (see also Simpson and Penrose \cite{SimpsonPenrose} and Hawking and Ellis \cite{HawkingEllis} as well as \cite{MisnerTaub})  one expects that the classical Einstein equations are such that, due to a blue-shift effect, such Cauchy horizons will be unstable against small perturbations, say of initial data in the relevant initially globally hyperbolic region and this leads one to expect/hope that the maximal Cauchy development of generic initial data close to that of exact Reissner Nordstr\"om or Kerr will be inextendible in a suitable sense  -- consistently with the strong version \cite{PenroseStrong} of Penrose's celebrated \emph{cosmic censorship conjecture}. (This may be taken to be the [somewhat vague] statement that the maximal Cauchy development of a generic initial data set should be a [globally hyperbolic] spacetime which is inextendible in a physically suitable sense.   For a recent survey of cosmic censorship in all its versions, see \cite{Landsman})  For what is known about this rather complicated and subtle question, we refer to the introduction to the paper \cite{DafShlapRough} of Dafermos and Shlapentokh-Rothman, the bottom line being, as one may read there, that, satisfactorily, such an inextendibility result does hold, in a particular sense proposed by Christodoulou \cite{ChrisDou}, which entails that the spacetime cannot be extended as a weak solution to Einstein's equations.\footnote{Christodoulou's proposal is to declare a spacetime to be extendible if it is extendible as a manifold and a chart may be found around each point of the extended manifold such that the metric components have continuous extensions and the Christoffel symbols are locally square integrable.}  Relatedly, and, say, taking Equation (\ref{kg}) as a proxy for more general perturbations, if one does QFT for Equation (\ref{kg}) on such spacetimes, then for a suitable notion of a general initial state, one expects the expectation value of the stress-energy tensor to diverge at the Cauchy horizon\footnote{This is stronger than just a failure of the two-point function of the initial state to extend as a Hadamard two-point function beyond the Cauchy horizon and corresponds to failure to be extendible in $H^1_\mathrm{loc}$.} sufficiently strongly as to lead one to expect that any solution to the semiclassical Einstein equations with such initial data would, again, be suitably inextendible beyond a maximal globally hyperbolic region.   Such a QFTCST divergence result has a long history -- let us just mention here the recent papers \cite{LanirEtAl} for Reissner-Nordstr\"om and \cite{OriOttPRL} for Kerr.  There is also other evidence that QFTCST effects cause Cauchy horizons to be unstable  from the study of the behaviour of gedanken particle detectors which move across such horizons.   See e.g.\ \cite{BeniParti,BeniJorma}. But in view of the classical results, we arguably don't need to rely on any sort of QFTCST results to conclude that strong cosmic censorship is saved for these spacetimes.

However, for Reissner Nordstr\"om-de Sitter and Kerr-de Sitter, the black hole blue shift effect (now for the Einstein equations with cosmological constant) is counteracted by a cosmological red-shift effect and, classically, it is known, after much work by many authors (see especially \cite{HintzVasy}) again for the `proxy' linear Equation (\ref{kg}) on such backgrounds that (at least for many values for the cosmological constant, the black hole mass and the scalar field mass) the singularity at the Cauchy horizon is too weak for the spacetime to be inextendible in the sense of Christodoulou -- thus seemingly insufficient to save strong cosmic censorship.

In the same paper \cite{DafShlapRough} by Dafermos and Shlapentokh-Rotman recommended above, a suggestion is made for a possible way out of this seeming impasse, staying within classical general relativity, by taking a suitable view as to which class of initial data is the right physical class -- in particular by taking a class of initial data which includes data that is suitably \emph{rough}.   However, it is of interest to explore whether quantum theory might save strong cosmic censorship, without having to get involved with debates about the right notion of roughness of classical initial data.   To this end, Hollands, Wald and Zahn \cite{HWZ} and Hollands, Klein and Zahn \cite{HKZ} have studied the expectation value of the quantum stress energy tensor again for the `proxy' quantum Equation (\ref{kg}) near the Cauchy horizon for all initially Hadamard states on Reissner Nordstr\"om-de Sitter spacetime and found, indeed that (generically in parameters -- see Footnote \ref{HWZgeneric}) this will blow-up as one approaches the Cauchy horizon strongly enough\footnote{\label{HWZgeneric} In suitable double-null Kruskal coordinates, $(U,V,\theta, \phi)$, adapated to the Cauchy horizon (see \cite{HWZ}) they show that, in any of their states and \emph{generically in parameters} by which we mean for almost all values of the cosmological constant, the black hole mass and charge and the scalar field mass, $\omega(T_{VV})$ blows up like $C/V^2$.} to lead one to expect that a solution to semiclassical gravity would 
convert the Cauchy horizon into a singularity through which the
spacetime could not be extended as a (weak) solution of the semiclassical Einstein equation and moreover be such that a physical object attempting to cross the horizon would be infinitely stretched or crushed.

We remark that an important step in the work in \cite{HWZ,HKZ} involves the construction of an Unruh-like state on a suitable region of Reissner-Nordstr\"om-de Sitter for massive or massless Klein Gordon.   (The r\^ole this plays is that then the two-point function in the generic initial Hadamard state they consider will differ from the two-point function in the Unruh state by a smooth two-point function.)   This has antecedents both in the work \cite{MarUnr} of  Markovi\'c and Unruh and also in the earlier work of Kay \cite{KayNantes} and Hollands \cite{HolPhD} towards a \emph{semi-local vacuum} notion for the neighbourhood of a point $p$ based on the imposition of a suitable vacuum-like condition on the boundary of the future light-cone of $p$.  In \cite{HWZ}, this semi-local vacuum construction is developed further and adapted to the region of Reissner-Nordstr\"om-de Sitter bounded in the past by the past black hole horizon and the past cosmological horizon.   Recently, Klein \cite{CKlein} has carried out an analogous construction of an Unruh-like state for Kerr-de Sitter -- an important step towards establishing the (expected) instability of the Kerr-de Sitter Cauchy horizon.

In conclusion, there seems now to be increasing evidence that, for de Sitter black holes, QFTCST has an important r\^ole to play in arguing that strong cosmic censorship is not violated.

\section{\label{NonGH} Non-Globally Hyperbolic Spacetimes and the ``Time Machine'' Question}

In (Hawking, 1992) it is argued that a spacetime in which a time-machine gets
manufactured should be modelled (see Figure 6) by a spacetime with an
{\it initial globally hyperbolic region} with a region containing {\it
closed timelike curves} to its future and such that the future boundary
of the globally hyperbolic region is a {\it compactly generated Cauchy
horizon} (see (Hawking, 1992) for the definition).   For such a spacetime, (Kay, Radzikowski and Wald, 1997) use microlocal methods and in particular certain propagation of singularity theorems to prove
that it is impossible for any distributional bisolution which
satisfies (even a certain weakened version of) the Hadamard condition on
the initial globally hyperbolic region to continue to satisfy that
condition on the full spacetime -- the (weakened) Hadamard condition
being necessarily violated at at least one point of the Cauchy horizon.
This result implies that, however one extends a state, satisfying our
conditions (a), (b), (c) and (d), on the minimal algebra for Equation
({\ref{kg}) on the initial globally hyperbolic region, the expectation
value of its stress-energy tensor must necessarily become singular, or rather undefinable, {\it
on} the Cauchy horizon.  This result, together with many heuristic
results and specific examples considered by many other authors was argued in (Kay, Radzikowski and Wald, 1997) to support the validity of the (Hawking, 1992) {\it chronology protection
conjecture} according to which it is impossible in principle to
manufacture a time machine.   For example, in (for simplicity, two-dimensional) Misner space\footnote{Misner space is a two-dimensional locally flat spacetime with the topology of a cylinder which is the disjoint union of a lower half-cylinder, which is  an initially globally hyperbolic region, an upper half-cylinder which has closed timelike curves and a compactly generated Cauchy horizon in between these two halves which consists of a null circle  -- see e.g.\ Hawking and Ellis \cite{HawkingEllis}.} one finds (see e.g.\ \cite{CramerKayTwo}) that, generically\footnote{\label{genparam} Note that here the genericity is in the state, in contrast to the situation for Reissner Nordstr\"om-de Sitter Cauchy horizons, where -- see Footnote \ref{HWZgeneric} in Section \ref{MoreBH}  -- it was in the parameters -- i.e.\ the cosmological constant, the black hole mass and charge and the field mass.} the expectation value of $T_{VV}$, where $V$ (see again \cite{CramerKayTwo}) is a coordinate\footnote{The $V$ coordinate can locally can be identified with the $V$ of standard double null coordinates, $(U,V)$, in Minkowski space such that the metric is $dUdV$.} that is regular across the Cauchy horizon, will behave like $C/V^2$ for a nonzero constant $C$.   However, the value of $C$ depends on the state, and, as pointed out by Krasnikov \cite{Krasnikov} there are states for which the value of $C$ is zero!  This is qualitatively different from the behaviour, discussed in Section \ref{MoreBH}, of the expectation value of the stress-energy tensor in initially Hadamard states near the Cauchy horizon of (1+3-dimensional) Reissner-Nordstr\"om-de Sitter where (see Footnote \ref{genparam}) as was shown in \cite{HWZ}, the constant, $C$, in the $C/V^2$ behaviour near the horizon (which holds except for certain discrete combinations of the parameters of that problem -- i.e.\ the cosmological constant, black hole mass and charge and field mass) is the \emph{same} for all initial Hadamard states!

The above-mentioned difference between the Reissner-Nordstr\"om-de Sitter Cauchy horizon story and the Misner-space Cauchy horizon story seems to help us to appreciate an important loophole, first argued for by Visser in \cite{VisserReliab} (see also (Visser, 2003)), in the physical interpretation of the Cauchy horizon quantum stress-energy result in the latter case.  Visser argues that the semiclassical approximation to quantum gravity might well be expected to be a bad approximation within a Planck distance or so (he gives a suitable definition for this notion in the above-cited papers) of the Cauchy horizon and yet, say in the Krasnikov state, the expectation value of the stress-energy tensor is only predicted to be singular/undefinable \emph{on} the Cauchy horizon, while it is bounded (in fact zero!) everywhere else (to the past of the Cauchy horizon).   Instead, Visser suggests that, before the Cauchy horizon is reached,  there may well be big quantum fluctuations in the geometry which would render any semiclassical approximation inapplicable and could only be understood in full quantum gravity and, for all we know, quantum gravity will effectively allow the Cauchy horizon to be traversed and thereby allow time travel to be possible and Hawking's Chronology Protection Conjecture to be violated.

To summarize: \emph{While the singularity/ill-definedness of the expectation value of the stress-energy tensor on the Cauchy horizon indicates a breakdown of semiclassical gravity there, that doesn't necessarily mean that physics breaks down there because semiclassical gravity may cease to be a good approximation to physics before one reaches the Cauchy horizon.  Therefore the question of whether or not chronology is protected is still open.}

\begin{figure}[h]
\label{fig:CGCH}
\centering
\includegraphics[trim = 4cm 20cm 5cm 4.5cm, clip]{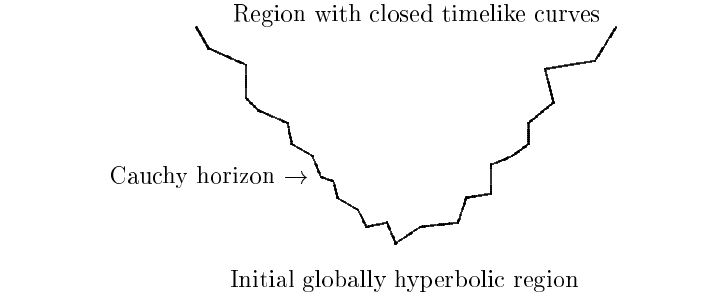}
\caption{\small The schematic geometry of a spacetime in which a
time-machine gets manufactured.}
\end{figure}

\section{\label{MoreNonGH} \textit{More about QFT on Non-Globally Hyperbolic Spacetimes}}

In 1992, Kay \cite{KayFlocality} proposed the \emph{F-locality} condition as a possible condition one might wish to impose on any net of local $*$-algebras on any (non-globally hyperbolic) spacetime.  Namely that every neighbourhood of every point should contain a globally hyperbolic subneighbourhood of that point for which the algebra coincides with the algebra one would obtain were one to regard the subneighbourhood as a spacetime in its own right and quantize -- with some choice of time-orientation -- according to the standard rules for quantum field theory on globally hyperbolic spacetimes.

Recently Janssen \cite{Janssen} has introduced an interesting general class of (not necessarily globally hyperbolic) spacetimes which he calls \emph{semi-globally hyperbolic spacetimes} and, for some of these, he has proposed a scheme for extending the usual construction of Equation (\ref{kg}) on globally hyperbolic spacetimes to spacetimes in this class with a resulting (minimal-like) net of local $*$-algebras which, inter alia, satisfies (a stronger condition than) the F-locality property.\footnote{Interestingly, Janssen arrives at this $*$-algebra construction by first constructing a patchwork of one-particle Hilbert space structures (see Footnote \ref{K}).  At first sight, this may seem strange since one traditionally thinks of a one-particle Hilbert space structure on a symplectic space of solutions as an auxiliary structure that helps to define certain (quasi-free) states on the $*$-algebra associated with that symplectic space.   Thus, in a sense, Janssen's construction gives primacy to states over field operators, in a reversal of the familiar story for globally hyperbolic spacetimes where the construction of a $*$-algebra takes primacy over the definition of states.}

Moreover, Janssen discusses a generalization of his construction to the (semi-globally hyperbolic) spacetime -- whose Penrose diagram first appeared in (Hawking, 1975) and is reproduced here in the diagram 7 --  which is assumed to be the best spacetime picture we can have, at a semiclassical level of description, for a (spherical) evaporating black hole. (A relevant recent paper with many references is \cite{MedaPinSiem}.)

\begin{figure}[h] 
\label{fig:evap}
\centering
\includegraphics[scale=0.7, trim = 4cm 17cm 5cm 4.5cm, clip]{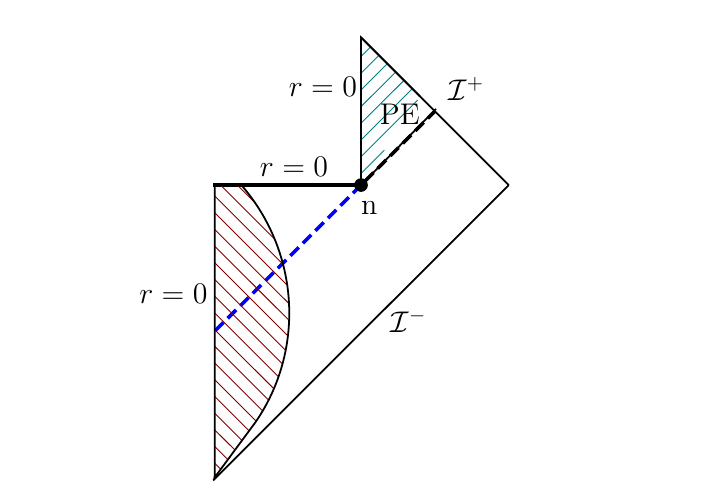}
\caption{\small The schematic geometry of a (spherically symmetric) spacetime of black-hole evaporation.  The backward sloping shaded region is the interior of the collapsing star, the dashed blue line is the black hole  horizon and the dashed black line (to the future of n) is the Cauchy horizon. PE is the post evaporation region.}
\end{figure}

Even though there are no closed timelike curves in this spacetime, it is nevertheless a non-globally hyperbolic spacetime since the point marked `n' on Figure 7 is a naked singularity (strictly this point is not in the spacetime manifold) and the dashed line to its future is a Cauchy horizon.   We shall call the region labelled `PE' in Figure 7 the \emph{post evaporation region}.  The complement of PE in this spacetime is easily seen to have Cauchy surfaces and therefore be globally hyperbolic; we shall call it the \emph{initial globally hyperbolic region}.  

Janssen obtains a net of local $*$-algebras for the QFT of Equation (\ref{kg}) on this spacetime.  However he leaves as an open question whether it admits any states.   As explained in his paper, there is a corresponding ``Weyl-like'' algebra which will admit states, but it seems to be unknown at present whether any of these states will satisfy the necessary regularity (see Footnote \ref{WeylReg} in Section \ref{staralgHad}) for any of their $n$-point functions to exist.  Thus one cannot yet even begin to talk about Hadamard states since of course the existence of states with $2$-point functions is a prerequisite for even saying what a Hadamard state would be!

The spacetime of Figure 7 is of importance since at least one version\footnote{The expression `black hole information loss puzzle' is also sometimes taken to refer to the distinct but closely related puzzle called in the most recent version of \cite{KayInfoLoss} `the second law puzzle in the case of black holes'.} of the \emph{black hole information loss puzzle} \cite{HawkingInfoLoss,UnruhWald,Maudlin,OkonSudarsky} is based on the seeming implication of this diagram that some of the information contained in an initial state on the initial globally hyperbolic region will be lost on the future singularity (the thick horizontal black line labelled $r=0$ in Figure 7) in the sense that the state on the initial globally hyperbolic region will not be retrodictable from knowledge of the state on the post-evaporation region in Figure 7.  And it is claimed to be difficult to imagine how this state of affairs could fail to persist in a suitable sense in full quantum gravity.

In a separate development, Ju\'arez-Aubry \cite{BeniConj}  has very recently put forward a number of general conjectures which entail the specific conjecture that, for any linear field theory (or interacting field theory when treated perturbatively) any extension of any pure Hadamard state on the initial globally hyperbolic region of the spacetime of Figure 7 to any F-local net of $*$-algebras for the full spacetime will fail to be Hadamard on the full spacetime.\footnote{A variation of this conjecture, which doesn't make any assumptions about the nature of any field algebra beyond the Cauchy horizon, and in particular therefore does not need to mention the F-locality condition, would be that, say for Equation (\ref{kg}), the two point function of any Hadamard state in the initial region will fail to have any Hadamard extension to beyond the Cauchy horizon.}  (As Ju\'arez-Aubry also points out, this conjecture, if true, is reminiscent of the result of Fewster and Verch in \cite{FewVerNecess} that a pure Hadamard state on a double cone in Minkowski space must fail to have a Hadamard extension beyond the double cone.)}  He further conjectures that, in fact, the state will fail to be Hadamard \emph{on} the Cauchy horizon sufficiently severely that the stress energy tensor will not be definable there -- rather as happens for the time-machine Cauchy horizons we discussed in Section \ref{NonGH}.

Ju\'arez-Aubry then points out that, if these conjectures hold, then one could conclude that semiclassical gravity would break down on the Cauchy horizon.  Thus it would not be true that the entire Figure 7 depicts a solution to semiclassical gravity.   To have a semiclassical solution, one would e.g.\ have to remove the PE region from Figure 7.  (See Figure 2 in \cite{BeniConj}.)  But there is no information loss puzzle associated with the resulting truncated figure!   So, assuming his conjectures are true, he would be able to say \emph{that it had actually never been true that any sort of information loss puzzle was ever suggested by any solution to semiclassical gravity appropriate to black hole evaporation.}

This is of course not to say that such a semiclassical solution, with the PE region excised, would be physically correct.  If it were, it would seem to mean that spacetime comes to an end at the (retarded) final moment of black hole evaporation! Rather, and somewhat similarly to Visser's conclusions in the case of the time-machine Cauchy horizons discussed in Section \ref{NonGH}, it seems reasonable to assume that quantum gravity would cause both the black hole singularity and the Cauchy horizon to be resolved and, insofar as it is representable by a classical spacetime picture, the solution to quantum gravity to again include a post-evaporation region.  But there is no reason to assume that this spacetime would resemble Figure 7.   Instead it is argued in \cite{BeniConj} that it is more reasonable to assume it would be globally hyperbolic (as depicted in Figure 3 in \cite{BeniConj}).  It is also pointed out in \cite{BeniConj} that the conjectured ill-definedness of the stress-energy tensor on the Cauchy horizon corresponds to the `thunderbolt' singularity argued for by Hawking and Stewart in \cite{HawkStew}.  Hawking and Stewart suggest  in that paper that, rather than spacetime coming to an end, quantum gravitational effects ``would soften the thunderbolt singularity into a burst of high energy particles'' at the endpoint of black hole evaporation.   Such a physical expectation would, as is pointed out in \cite{BeniConj} seem to be consistent with the globally hyperbolic spacetime depicted in Figure 3 in \cite{BeniConj}.

\section{\label{OtherWarn} Other Related Topics and Some Warnings}

There is a vast computational literature, calculating the expectation
values of stress-energy tensors in states of interest for scalar and
higher spin linear fields (and also some work for interacting fields) on
interesting cosmological and black-hole backgrounds.  For an early survey, see (Birrell and Davies, 1982).  For a few recent papers on computational and approximation methods, see e.g.\ 
\cite{AndEak,LevOri,TayBreeOtt} as well, e.g.\ as recent papers such as \cite{LanirEtAl,HWZ,HKZ} where (see Section \ref{MoreBH}) a computation of an expectation value of a renormalized stress-energy tensor forms forms part of a specific investigation.

Quantum field theory on de Sitter space is a big subject area in
its own right with many interesting features, while quantum field theory on anti-de Sitter space is of sustained interest because of its relevance to
{\it string theory} and {\it holography}.   Also important on black hole
backgrounds is the calculation of grey-body factors, again with renewed
interest because of relevance to string theory and to {\it brane-world} scenarios.

Further to the topics discussed in Section \ref{AQFT}, there are many further mathematically rigorous results on algebraic and axiomatic quantum field theory in a curved spacetime setting, including versions of the {\it PCT} (see \cite{HollandsPCT}), {\it Spin-Statistics} (see Verch's paper, \cite{VerchSpinStat}, which we mentioned above and also \cite{FewsterSpinStat})  {\it Reeh-Schlieder} theorems (see \cite{StrohEtAl2002,KoSanReehSchl} and also the very recent paper \cite{StrohWitten} which also generalizes certain theorems of Araki and Borchers) as well as the \emph{split property} \cite{FewSplit1,FewSplit2}.

There are also rigorous {\it energy inequalities} bounding the extent
to which expected energy densities etc.\ can be negative. For more about this extensive topic see \cite{FewstLectEnIn,KonSanEnInequ}.  See also \cite{WittenEnIneq}.

Perturbative renormalization theory of interacting QFT (including gauge theories -- see e.g.\ \cite{HollYM}) in
curved spacetime is also now a highly developed subject, which, in the context of AQFT, is known as {\it pAQFT}.   For more information about pAQFT, see the article \cite{FredRejArt} by Fredenhagen and Rejzner and the book \cite{RejznerBook} by Rejzner as well as the article \cite{RejznerEncyc} by Brunetti, Fredenhagen and Rejzner in this encyclopedia.  pAQFT not only generalises what was previously understood about perturbative renormalization in Minkowski space to curved spacetimes but has also achieved a new understanding (see e.g.\ the book \cite{DutschBook} by D\"utsch) as to how pAQFT can be understood as arising from a \emph{deformation} of a perturbative treatment of classical field theory.   Moreover (see \cite{FredRejBack}) it has been shown how pAQFT when combined with ideas from locally covariant QFT (see Section \ref{AQFT}) leads to a new understanding of how perturbative QFT can be consistent with background independence and this, in turn, has led (see \cite{BrunFredRej,BrunFredHackPinRej}) to a new background-independent perturbative approach to quantum gravity\footnote{As pointed out in \cite{BrunFredHackPinRej}, this work in some respects supercedes other approaches to (classical and quantum, gauge invariant, nonlinear) cosmological perturbation theory.}, which is also known as a \emph{relational} (and gauge-invariant) version of perturbation theory.   For an illustration of its power, and fuller references to earlier work, see e.g.\ the recent papers \cite{Lima, FrobReinVerch} by Lima and by Fr\"ob, Rein and Verch. Treating perturbative quantum gravity as an effective field theory, the former computes quantum-gravitational one-loop corrections to a gauge-invariant observable measuring the space-time local expansion rate in slow-roll inflation thereby contributing to the resolution of some earlier controversies and ambiguities; the latter computes quantum gravitational corrections to the Newtonian potential, giving, for the first time, unambiguous answers for the appropriate generalization to non-trivial backgrounds. (See also \cite{BrunFredRejEff}.)

Let us also mention here the construction of scalar quantum fields with polynomial interactions in 2-dimensional de Sitter space by Barata, J\"akel and Mund -- see \cite{BJM} and references therein.   As one may read there, this work makes interesting and novel use of the notion of one-particle structures (see Footnote \ref{K}) and of Tomita-Takesaki modular theory and also introduces and uses a new version of the Osterwalder-Schrader reconstruction theorem.  

Beyond quantum field theory in a fixed curved spacetime is {\it
semiclassical gravity} which takes into account the {\it back reaction}
of the expectation value of the stress-energy tensor on the classical
gravitational background and which we have referred to in several places in this article.  It is beyond the scope of this article to review the subject.  But we would mention the review article, \cite{Ford} of Ford and the book \cite{HuVerd} of Hu and Verdaguer.   Aside from \cite{JuaKayMirSud} which we referred to in Section \ref{MoreHad}, an incomplete, but we hope useful, list of recent references is \cite{BeniCosConst,BeniTonDan,GottSiem,MedPinSiem,
GottRothSiem1,KoSanSemi,BeniSemiConstr,BeniModak,MedPin,GottRothSiem2}.

Readers exploring the wider literature, or doing further research on the
subject should be  aware that the word ``vacuum'' is sometimes used to
mean ``ground state'' and sometimes just to mean (pure) ``quasi-free state''. 
Furthermore they should be cautious of attempts to define
particles on Cauchy surfaces in {\it instantaneous diagonalization}
schemes (cf.\ the remarks at the end of Section \ref{particle}).   When studying (or
performing) calculations of the ``expectation value of the stress-energy
tensor'' it is always important to ask oneself with respect to which
state the expectation value is being taken.  It is also important to
remember to check that candidate two-point (anticommutator) functions
satisfy the positivity condition (c) of Section \ref{staralgHad}.  Typically two-point
distributions obtained via {\it mode sums} automatically satisfy
Condition (c) (and Condition (d)), but those obtained via {\it image}
methods don't always satisfy it.  (When they don't, the presence of
nonlocal spacelike singularities is often a tell-tale sign as can be
inferred from Kay's Conjecture/Radzikowski's Theorem discussed in Section
\ref{staralgHad}.) There are a number of apparent implicit assertions in the literature
that some such two-point functions arise from ``states'' when of course
they can't.  Some of these concern proposed analogues to the
Hartle-Hawking-Israel state for the (appropriate maximal globally hyperbolic
portion of the maximally extended) Kerr spacetime.  That they can't
belong to states is clear from the theorem in (Kay and Wald, 1991) (see Section \ref{blackholes}) which
states that there is no isometry-invariant Hadamard state on this spacetime at
all.   Others of them concern claimed ``states'' on spacetimes such as
those discussed in Section \ref{NonGH} which, if they really were states, would
seem to be in conflict with the chronology protection conjecture. (In the absence of a full understanding of quantum gravity, it is of course still unknown whether or not that conjecture holds, but, as far as I am aware, there are still no known violations of it which are semiclassically describable.)  Finally, beware states  (such as the so-called {\it $\alpha$-vacua} of de Sitter spacetime) whose two-point distributions violate the ``Hadamard'' Condition (d) of Section 2 and which therefore do not have
a well-defined finite expectation value for the renormalized
stress-energy tensor.   As we discussed in Section \ref{MoreHad} the same is true of the (unmodified) S-J states.

\section*{See also}

\textbf{Algebraic approach to quantum field theory. The operator product expansion in quantum field theory. Measurement in quantum field theory. Perturbative algebraic quantum field theory and beyond. Scattering in relativistic quantum field theory: basic concepts, tools, and results.}

\section*{Acknowledgments} I thank Chris Fewster, Atsushi Higuchi and Benito Ju\'arez-Aubry for many suggestions and Klaas Landsman for commenting on an earlier version.

\section*{\label{FR} Further Reading}

\begin{description}

\item DeWitt BS (1975) Quantum field theory in curved space-time.
{\it Physics Reports} 19, No. 6, 295-357.

\item Birrell ND and Davies PCW (1982) {\it Quantum Fields in Curved Space.}
Cambridge: Cambridge University Press.

\item Haag R (2012) {\it Local Quantum Physics.}  Berlin: Springer.

\item Dimock J (1980) Algebras of local observables on a manifold. {\it
Communications in Mathematical Physics} 77: 219-228.

\item Brunetti R, Fredenhagen K and Verch R (2003)
The generally covariant locality principle -- A new paradigm for local quantum
physics.
{\it Communications in Mathematical Physics} 237:31-68.

\item Kay, BS (2000) Application of linear hyperbolic PDE to linear quantum
fields in curved spacetimes: especially black holes, time machines and a
new semi-local vacuum concept.  Journ\'ees {\it \'Equations aux
D\'eriv\'ees Partielles},  Nantes, 5-9 juin 2000 GDR 1151 (CNRS): IX1-IX19.
(Also available at http://www.math.sciences.univ-nantes.fr/edpa/2000/html
or as gr-qc/0103056.)

\item Wald RM (1978) Trace anomaly of a conformally invariant
quantum field in a curved spacetime.  {\it Physical Review} D17: 1477-1484.

\item Kay BS (1979) Casimir effect in quantum field theory. ({\it
Original title:} The Casimir effect without magic.) {\it
Physical Review} D20: 3052-3062.

\item Hawking SW (1975) Particle Creation by Black Holes.  {\it
Communications in Mathematical Physics} 43: 199-220.

\item Kay BS and Wald RM (1991) Theorems on the uniqueness and
thermal properties of stationary, nonsingular, quasifree states on
spacetimes with a bifurcate Killing horizon. {\it Physics Reports} 207,
No. 2, 49-136.

\item Kay BS (1993) Sufficient conditions for quasifree states and an
improved uniqueness theorem for quantum fields on space-times with
horizons. {\it Journal of Mathematical Physics} 34: 4519-4539.

\item Wald RM (1994) {\it Quantum Field Theory in Curved Spacetime and Black
Hole Thermodynamics.} Chicago: University of Chicago Press.

\item Hartle JB and Hawking SW (1976) Path-integral derivation of
black-hole radiance.  {\it Physical Review} D13: 2188-2203.

\item Israel W (1976) Thermo-field dynamics of black holes.
{\it Physics Letters A} 57: 107-110.

\item Unruh W (1976) Notes on black hole evaporation.  {\it Physical
Review} D14: 870-892.

\item Hawking SW (1992) The Chronology Protection Conjecture.
{\it Physical Review} D46: 603-611.

\item Kay BS, Radzikowski MJ and Wald RM (1997) Quantum field theory
on spacetimes with a compactly generated Cauchy horizon.  {\it
Communications in Mathematical Physics} 183: 533-556.

\item Visser M (2003) The quantum physics of chronology protection.  In:
Gibbons GW, Shellard EPS and Rankin SJ {\it The Future of Theoretical Physics
and Cosmology.} Cambridge: Cambridge University Press.

\item Parker L and Toms D (2009) {\it Quantum Field Theory In Curved Spacetime} Cambridge: Cambridge University Press.

\item Mukhanov V and Winitzki S (2007) {\it Introduction to Quantum Effects in Gravity.} Cambridge: Cambridge University Press.

\item Brunetti R, Dappiaggi C, Fredenhagen K and Yngvason J eds.\ (2015) {\it Advances in Algebraic Quantum Field Theory.} Cham: Springer International Publishing. \url{https://link.springer.com/book/10.1007/978-3-319-21353-8}

\end{description}


\begin{thebibliography}{999}


\bibitem{encyc2006} Kay BS (2006) Quantum field theory in curved spacetime.  In: 
Fran\c coise J-P, Naber G and Tsou TS (eds) {\it Encyclopedia of Mathematical Physics.}
Amsterdam, New York, London: Academic (Elsevier) 4: 202-212. [arXiv:gr-qc/0601008]

\bibitem{BarCamFuenI} Barbado LC, B\'aez-Camargo and Fuentes I (2021): Evolution of confined quantum scalar fields in curved spacetime. Part II: Spacetimes with moving boundaries in any synchronous gauge.
{\it European Physics Journal C} 81: 953. [arXiv:1811.10507]

\bibitem{BarCamFuenII} Barbado LC, B\'aez-Camargo and Fuentes I  (2020) Evolution of confined quantum scalar fields in curved spacetime. Part I: Spacetimes without boundaries or with static boundaries in a synchronous gauge.  {\it European Physical Journal C} 80: 796. [arXiv:2106.14923]

\bibitem{KohLouFueBru} Kohlrus J, Louko J, Fuentes I and Bruschi DE (2018): Wigner phase of photonic helicity states in the spacetime of the Earth. arXiv:1810.10502.

\bibitem{UnruhDumb} Unruh WG (2008) Dumb holes: analogues for black holes. {\it Philosophical Transactions of the Royal Society A: Mathematical, Physical and Engineering Sciences} 366: 2905-2913.

\bibitem{BarLibVis} Barcelo C, Liberati S and Visser M (2011) Analogue Gravity. {\it Living Reviews in Relativity} 14: 3. \url{http://dx.doi.org/10.12942/lrr-2011-3}

\bibitem{WeinTedPenUnrLaw} Weinfurtner S, Tedford EW, Penrice MCJ, Unruh WG, Lawrence GA (2011)
Measurement of stimulated Hawking emission in an analogue system.
{\it Phys.~Rev.~Lett.} 106: 021302. [arXiv:1008.1911]

\bibitem{RosBerSilLeo} Drori J, Rosenberg Y, Bermudez D, Silberberg Y and Leonhardt U (2019) Observation of Stimulated Hawking Radiation in an Optical Analogue. {\it Phys.~Rev.~Lett.} 122: 010404. 

\bibitem{MerMogCheMorBriWez} Mertens L, Moghaddam AG, Chernyavsky D, Morice C, van den Brink J and van Wezel J (2002)
Thermalization by a synthetic horizon. {\it Phys.~Rev.~Research} 4: 043084. 
[arXiv:2206.08041]

\bibitem{BrauFaiEtAl} Braunstein SL, Faizal M, Krauss LM, Marino F and Shah NA (2023) Analogue simulations of quantum gravity with fluids. {\it Nat.~Rev.~Phys.} (2023) \url{https://doi.org/10.1038/s42254-023-00630-y}

\bibitem{WittenQFTCST} Witten E (2022) Why does quantum field theory in curved spacetime make sense? And what happens to the algebra of observables in the thermodynamic limit? In: {\it Dialogues Between Physics and Mathematics: CN Yang at 100.} (Cham: Springer International Publishing) pp.~241-284. [arXiv:2112.11614]

\bibitem{Gibbons79} Gibbons GW (1979) Quantum field theory in curved spacetime.  In: Hawking SW, Israel W (eds) (1979) {\it General Relativity, An Einstein Centenary Survey}  Cambridge, Cambridge University Press.  

\bibitem{Ford97} Ford LH (1997) Quantum field theory in curved spacetime. arXiv: gr-qc/9707062

\bibitem{Jacobson2005}  Jacobson T (2005) Introduction to quantum fields in curved spacetime and the Hawking effect.  arXiv:gr-qc/0308048

\bibitem{WaldHist12} Wald RM (2012) The history and present status of quantum field theory in curved spacetime  In: Lehner C, Renn J and Schemmel M (eds) {\it Einstein and the Changing Worldviews of Physics. Einstein Studies, vol 12} Birkh\"auser Boston. pp. 317-331. [arXiv:gr-qc/0608018]

\bibitem{HollWaldPhysRep} Hollands S and Wald RM (2015) Quantum fields in curved spacetime. {\it Physics Reports} 574: 1-35.  [arXiv:1401.2026]

\bibitem{WaldFormul} Wald RM (2018) The formulation of quantum field theory in curved spacetime. In: Rowe DE, Sauer T, Walter SA (eds) {\it Beyond Einstein: Perspectives on Geometry, Gravitation, and Cosmology in the Twentieth Century} New York, Birkh\"auser.  pp.~439-449.  [arXiv:0907.0416]

\bibitem{BrunFred} Brunetti R and Fredenhagen K (2009) Quantum field theory on curved backgrounds.  In: B\"ar C and Fredenhagen K (eds) (2009) {\it Quantum Field theory on Curved Spacetimes: Concepts and Mathematical Foundations} Berlin: Springer [arXiv:0901.2063]

\bibitem{BaerGP} B\"ar C, Ginoux N and Pf\"affle F (2007) {\it Wave Equations on Lorentzian Manifolds and Quantization} European Mathematical Society. [arXiv:0806.1036]

\bibitem{FewsterVerch2015} Fewster CJ and Verch R (2015) 
Algebraic quantum field theory in curved spacetimes.  In: (Brunetti et al., 2015) (see Further Reading) 125-189.
[arXiv:1504.00586]

\bibitem{Hack} Hack TP (2015) {\it Cosmological applications of algebraic quantum field theory in curved spacetimes} Cham: Springer. [arXiv:1506.01869]

\bibitem{HollandsEncyc} Hollands S and Wald RM (2025) The Operator Product Expansion in Quantum Field Theory. Encyclopedia of Mathematical Physics 5, 382-391. [arXiv:2312.01096]

\bibitem{MorettiCorrn} Moretti V (2021) On the global Hadamard parametrix in QFT and the signed squared geodesic distance defined in domains larger than convex normal neighbourhoods. {\it Letters in Mathematical Physics} 111: 130. [arXiv:2107.04903]

\bibitem{DecFol} D\'ecanini Y and Folacci A (2008) Hadamard renormalization of the stress-energy tensor for a quantized scalar field in a general spacetime of arbitrary dimension. {\it Physical Review D} 78: 044025. [arXiv:gr-qc/0512118]

\bibitem{FewsterVerch2012a} Fewster CJ and Verch R (2012) Dynamical locality and covariance: What makes a physical theory the same in all spacetimes? {\it Annales Henri Poincar\'e} 13: 1613-1674. [arXiv:1106.4785]

\bibitem{Hartman} Hartman T (2015) Lectures on Quantum Gravity and Black Holes. (Cornell University)
\url{http://www.hartmanhep.net/topics2015/gravity-lectures.pdf}

\bibitem{BeniConj} Ju\'arez-Aubry BA (2023)  Quantum strong cosmic censorship and black hole evaporation.  arXiv:2305.01617.

\bibitem{MTW} Misner, C. W., Thorne, K. S. and Wheeler, J. A. (1973) Gravitation. San Francisco: Freeman.

\bibitem{BerSan1} Bernal AN and S\'anchez M (2003) On smooth Cauchy hypersurfaces and Geroch's splitting theorem.  {\it Commun.~Math.~Phys.} 243: 461-470.

\bibitem{FewRej} Fewster CJ and Rejzner K (2020) Algebraic quantum field theory: An introduction. In: Finster F, Giulini D, Kleiner J and Tolksdorf J (eds)\ {\it Progress and Visions in Quantum Theory in View of Gravity: Bridging Foundations of Physics and Mathematics.}  Cham, Birkh\"auser. pp.\ 1-61.  \url{https://link.springer.com/chapter/10.1007/978-3-030-38941-3_1} [arXiv:1904.04051]

\bibitem{BuchFred} Buchholz D and Fredenhagen K (2025) Algebraic quantum field theory: objectives, methods, and results. Encyclopedia of Mathematical Physics 5, 278-290. [arXiv:2305.12923]

\bibitem{KayDoubleWedge} Kay BS (1985) The double-wedge algebra for quantum fields on Schwarzschild and Minkowski spacetimes. {\it Communications in Mathematical Physics} 100: 57-81.  And Erratum to appear.

\bibitem{RadzKayConj} Radzikowski MJ (1996): A Local-to-Global Singularity Theorem for
Quantum Field Theory on Curved Space-Time. {\it Commun.~Math.~Phys.} 180: 1-22.  (Includes an Appendix by Verch R.)

\bibitem{KayThesis} Kay BS (1977) Quantum fields in time-dependent backgrounds and in curved spacetimes.
PhD thesis, University of London. 

\bibitem{KaySeg} Kay BS (1979) A uniqueness result in the Segal-Weinless approach to linear Bose fields. {\it Journal of Mathematical Physics} 20:1712-1713.

\bibitem{KayCMP78} Kay BS (1978) Linear spin-zero quantum fields in external gravitational and scalar fields. {\it Commun.~Math.~Phys} 62: 55-70.

\bibitem{KayKMS} Kay BS (1985) A uniqueness result for quasi-free KMS states. (1985) {Helv.~Phys.~Acta} 58:1017-1029.

\bibitem{RadzMicro} Radzikowski MJ (1996): Micro-local approach to the Hadamard condition in quantum field theory on curved space-time.  {\it Commun.~Math.~Phys.} 179: 529-553. 

\bibitem{FewWorLin2000} Fewster CJ (2000) A general worldline quantum inequality. 
{\it Class.~Quantum~Grav.} 17:1897. [arXiv:gr-qc/9910060]

\bibitem{FewVerNecess} Fewster CJ and Verch R (2013) The Necessity of the Hadamard Condition. {\it Class. Quant. Grav.} 30:235027. [arXiv:1307.5242]

\bibitem{KurpiczEtAl} Kurpicz F, Pinamonti N and Verch R (2021) Temperature and entropy-area relation of quantum matter near spherically symmetric outer trapping horizons. {\it Letters in Mathematical Physics}: 111: 110. [arXiv:2102.11547]

\bibitem{SahlVerch} Sahlmann H and Verch R (2000) Passivity and microlocal spectrum condition. { \it Commun.~ Math.~Phys.} 214: 705. [arXiv:math-ph/0002021]
 
\bibitem{KoSan2013} Sanders K  (2013) Thermal equilibrium states of a linear scalar quantum field in stationary spacetimes. {\it International Journal of Modern Physics A} 28: 1330010. [arXiv:1209.6068]


\bibitem{VerchConjProof} Verch R (1994) Local definiteness, primarity and quasiequivalence of quasifree Hadamard quantum states in curved spacetime. {\it Commun.~Math.~Phys.} 160: 507-536.


\bibitem{SanHad} Sanders K (2010) Equivalence of the (generalised) Hadamard and microlocal spectrum condition for (generalised) free fields in curved spacetime. Commun.~Math.~Phys. 295: 485–501. [arXiv:0903.1021]

\bibitem{KhavMor} Khavkine I and Moretti V (2015) Algebraic QFT in curved spacetime and quasifree Hadamard states: an introduction. In: (Brunetti et al., 2015) (see Further Reading) 191-251.  [arXiv:1412.5945]

\bibitem{FullSweeWald} Fulling SA, Sweeny M and Wald RM (1978) Singularity structure of the two-point function in quantum field theory in curved spacetime. {\it Communications in Mathematical Physics} 63: 257-264. 

\bibitem{FullNarcoWald} Fulling SA, Narcowich FJ and Wald RM (1981) Singularity structure of the two-point function in quantum field theory in curved spacetime, II. {\it Annals of Physics} 136: 243-272.

\bibitem{JunkerRev} Junker W (1996) 
Hadamard States, Adiabatic vacua and the construction of physical states for scalar quantum fields on curved spacetime. {\it Reviews in Mathematical Physics} 08: 1091-1159.   And Erratum (2002)
{\it Reviews in Mathematical Physics} 14: 511-51.

\bibitem{GerWrochPseud} G\'erard C and Wrochna M (2014) Construction of Hadamard states by pseudo-differential calculus. {\it Commun.~Math.~Phys.} 325: 713-755. [arXiv:1209.2604]

\bibitem{GerWrochCald} G\'erard C and Wrochna M (2019) Analytic Hadamard States, Calder\'on Projectors and Wick Rotation Near Analytic Cauchy Surfaces. {\it Commun.~Math.~Phys.} 366: 29-65. [arXiv:1706.08942]

\bibitem{GerWrochInOut}  G\'erard C and Wrochna M (2017) Hadamard property of the in and out states for Klein-Gordon fields on asymptotically static spacetimes. {\it Annales Henri Poincar\'e} 18: 2715-2756. 

\bibitem{MoMuVo} Moretti V, Murro S and Volpe D (2023) Paracausal deformations of Lorentzian metrics and M\o ller isomorphisms in algebraic quantum field theory. Sel.\ Math.\ New Ser.\ 29, 56  [arXiv:2109.06685]

\bibitem{DapMorPinHadLight} Dappiaggi C, Moretti V, Pinamonti N (2017) Hadamard States From Light-like Hypersurfaces {\it SpringerBriefs in Mathematical Physics 2017} [arXiv:1706.09666]

\bibitem{GerWrochChar} G\'erard C and Wrochna M (2016) Construction of Hadamard states by characteristic Cauchy problem. {\it Analysis and PDE} 9: 111-149. [arXiv:1409.6691]

\bibitem{HollDirac} Hollands S (2001) The Hadamard condition for Dirac fields and adiabatic states on Robertson-Walker spacetimes. {\it Communications in Mathematical Physics} 216: 635-661. [arXiv:gr-qc/9906076]

\bibitem{GerStosk} G\'erard C and Stoskopf T (2022) Hadamard states for quantized Dirac fields on Lorentzian manifolds of bounded geometry. {\it Reviews in Mathematical Physics} 34: 2250008. [arXiv:2108.11630]

\bibitem{FewPfen} Fewster CJ, Pfenning MJ (2003). A quantum weak energy inequality for spin-one fields in curved space-time. {\it Journal of Mathematical Physics} 44: 4480-4513.  [arXiv:gr-qc/0303106]

\bibitem{WrochnaZahn} Wrochna M and Zahn J (2017) Classical phase space and Hadamard states in the BRST formalism for gauge field theories on curved spacetime. {\it Reviews in Mathematical Physics} 29: 1750014. [arXiv:1407.8079]

\bibitem{GerardWrochnaYM} G\'erard C and Wrochna M (2015) Hadamard states for the linearized Yang-Mills equation on curved spacetime. {\it Communications in Mathematical Physics} 337: 253-320. [arXiv:1403.7153]


\bibitem{GerardMurroWrochnaLinGR} G\'erard C, Murro S and Wrochna M (2025) Wick Rotation of Linearized Gravity in Gaussian Time and Calder\'on Projectors.  Annales Henri Poincar\'e \url{https://doi.org/10.1007/s00023-025-01539-1} [arXiv:2204.01094]


\bibitem{MomMuVo2} Moretti V, Murro S and Volpe D (2023) The quantization of Proca fields on globally hyperbolic spacetimes: Hadamard states and M\o ller operators. Ann.~Henri~Poincar\'e 24: 3055-3111. [arXiv:2210.09278] 

\bibitem{FewPolar} Fewster CJ (2025) Polarisation sets of Green operators for normally hyperbolic equations.  arXiv:2503.12544

\bibitem {FewGreen} Fewster CJ (2025) Hadamard states for decomposable Green-hyperbolic operators. arXiv:2503.12537

\bibitem{BrumFred} Brum M and Fredenhagen K (2013) ‘Vacuum-like’ Hadamard states for quantum fields on curved spacetimes. {\it Classical and Quantum Gravity} 31: 025024. [arXiv:1307.0482]

\bibitem{FewArtState} Fewster CJ (2018) The art of the state. {\it International Journal of Modern Physics D} 27: 1843007. [arXiv:1803.06836]

\bibitem{SorkJohnst} Afshordi N, Aslanbeigi S and Sorkin RD (2012) A distinguished vacuum state for a quantum field in a curved spacetime: formalism, features, and cosmology. {\it Journal of High Energy Physics} 2012: 1-29. [arXiv:1205.1296]

\bibitem{FVRecentConstr} Fewster CJ and Verch R (2012)  On a recent construction of ‘vacuum-like’ quantum field states in curved spacetime. {Class.~Quantum~Grav.} 29: 205017. [arXiv:1206.1562]

\bibitem{JuaKayMirSud} Ju\'arez-Aubry BA, Kay BS, Miramontes T, Sudarsky D (2023) On the initial value problem for semiclassical gravity without and with quantum state collapses. {\it Journal of Cosmology and Astroparticle Physics} 01: 040. [arXiv:2205.11671]

\bibitem{BerSan2} Bernal AN and S\'anchez M (2005) Smoothness of time functions and the metric splitting
of globally hyperbolic spacetimes.  {\it Commun.~Math.~Phys.} 257: 43-50. [arXiv:gr-qc/0401112]

\bibitem{Shale} Shale D (1962) Linear symmetries of free Boson fields.   {\it Transactions of the American Mathematical Society} 103: 149-167.

\bibitem{Wald79} Wald RM (1979) Existence of the S-matrix in quantum field theory in curved space-time. {\it Annals of Physics} 118: 490-510.

\bibitem{Dimock79} Dimock J (1979) Scalar quantum field in an external gravitational field. {\it Journal of Mathematical Physics} 20:  2549-2555.

\bibitem{Kay82} Kay BS (1982)  Quantum fields in curved space-times and scattering theory. In: Doebner HD, Andersson SI, Petry HR (eds) (1980) {\it Differential Geometric Methods in Mathematical Physics: Proceedings, Clausthal}  Springer Lecture Notes in Mathematics 905 pp.\ 272-295. Berlin-Heidelberg: Springer.

\bibitem{DimockKaySchwI} Dimock J and Kay BS (1987) Classical and quantum scattering theory for linear scalar fields on the Schwarzschild metric I.{\it  Annals of Physics} 175: 366-426.

\bibitem{DucDyb} Duch P and Dybalski W (2023) Infrared problem in quantum electrodynamics. Encyclopedia of Mathematical Physics 5, 304-316. [arXiv:2307.06114]

\bibitem{MorettiCommStress} Moretti V (2003) Comments on the stress-energy tensor operator in curved spacetime.  {\it Commun.~Math.~Phys.} 232:189-221. [arXiv:gr-qc/0109048]

\bibitem{BalaWin} Balakumar V and Winstanley E (2020) Hadamard renormalization for a charged scalar field. {\it Classical and Quantum Gravity} 37: 065004.

\bibitem{Verch1999} Verch R (1999) Wavefront sets in algebraic quantum field theory. {\it Commun.\ Math.\ Phys.} 205: 337-367. [arXiv:math-ph/9807022]

\bibitem{BrunFred2000} Brunetti R and Fredenhagen K (2000) Microlocal analysis and
interacting quantum field theories: Renormalization on physical backgrounds.  {\it Communications in Mathematical Physics} 208: 623-661. [arXiv:math-ph/9903028]

\bibitem{FewsterVerch2012b} Fewster CJ and Verch R (2012) Dynamical locality of the free scalar field. {\it Annales Henri Poincar\'e} 13: 1675-1709. [arXiv:1109.6732]

\bibitem{FredHaag} Fredenhagen K and Haag R (1990) On the derivation of Hawking radiation associated with the formation of a black hole. {\it Communications in mathematical physics} 127: 273-284.

\bibitem{HawkingEllis} Hawking SW and Ellis GFR (1973) {\it The large scale structure of space-time.} Cambridge: Cambridge University Press.

\bibitem{WaldTrans} Wald RM (1976) Stimulated-emission effects in particle creation near black holes.  {\it Physical Review D} 13: 3176. 

\bibitem{LeahyUnruh} Leahy DA and Unruh WG (1983). Effects of a $\lambda\Phi^4$ interaction on black-hole evaporation in two dimensions. {\it Physical Review D} 28: 694.

\bibitem{BachAsymp} Bachelot A. (1994) Asymptotic completeness for the Klein-Gordon
equation on the Schwarzschild metric.
{Annales de l'Institut Henri Poincar\'e section A} 61: 411-441.

\bibitem{DimAsymp} Dimock J (1985) Scattering for the Wave Equation on the Schwarzschild Metric.  
{Gen.~Rel.~Grav.} 17: 353-369.

\bibitem{BachHawk} Bachelot A (1999) The Hawking effect. {\it Annales de l'Institut Henri Poincar\'e, Physique th\'eorique} 70: 41-99.

\bibitem{Melnyk} Melnyk F (2004) The Hawking Effect for Spin 1/2 Fields.  {\it Commun.~Math.~Phys.} 244: 483-525 

\bibitem{AlfordPhD} Alford F (2022) A Mathematical Study of Hawking Radiation on Collapsing, Spherically Symmetric Spacetimes. (Doctoral dissertation, University of Cambridge). \url{https://api.repository.cam.ac.uk/server/api/core/bitstreams/8e4ec008-0f5d-4711-b78e-67bf01aff0d3/content}

\bibitem{AlfordPaper1} Alford, F (2020) The scattering map on Oppenheimer-Snyder space-time.  {\it Annales Henri Poincar\'e} 21: 2031-2092. [arXiv:1910.02752]

\bibitem{AlfordPaper2} Alford, F (2023) The scattering map on collapsing charged spherically symmetric spacetimes. arXiv:2309.03022

\bibitem{AlfordPaper3} Alford, F (2023) A rigorous study of Hawking radiation on collapsing charged spherically symmetric spacetimes. arXiv:2309.03022

\bibitem{DafRodLect} Dafermos M and Rodnianski I (2013) Lectures on black holes and linear waves. {\it Clay~Math.~Proc} 17: 97-205. [arXiv:0811.0354]

\bibitem{DafRodShl} Dafermos M, Rodnianski I and Shlapentokh-Rothman Y (2018)
A scattering theory for the wave equation on Kerr black hole exteriors.
{\it Annales Scientifiques de l'ENS}
51: 371-486 [arXiv:1412.8379]

\bibitem{HafNicScatt} H\"afner D and Nicolas JP (2004) Scattering of massless Dirac fields by a Kerr black hole. {\it Reviews in Mathematical Physics} 16: 29-123.

\bibitem{Batic} Batic D (2007) Scattering for massive Dirac fields on the Kerr metric. {\it Journal of mathematical physics} 48: 022502. [arXiv: gr-qc/0606051]

\bibitem{HiguchiYamamoto2018} Higuchi A and Yamamoto (2018) Vacuum state in de Sitter spacetime with static charts. {\it Phys.\ Rev.} D98: 065014. [arXiv:1808.02147]

\bibitem{CrispinoEtAl} Crispino LCB, Higuchi A and Matsas GEA (2008) The Unruh effect and its applications. {\it Rev.\ Mod.\ Phys.} 80: 787. [arXiv:0710.5373]

\bibitem{Sewell80} Sewell GL (1980) Relativity of temperature and the Hawking effect. {\it Physics Letters A} 79: 23-24.

\bibitem{Sewell82} Sewell GL (1982) Quantum fields on manifolds: PCT and gravitationally induced thermal states. {\it Annals of Physics} 141: 201-224.

\bibitem{Jacobson94} Jacobson T (1994) A note on Hartle-Hawking vacua  {\it Phys.\ Rev.} D50:6031-6032. [arXiv: gr-qc/9407022]

\bibitem{SandersHHI} Sanders K (2015) On the construction of Hartle-Hawking-Israel
states across a static bifurcate Killing horizon.  {Lett.~Math.~Phys.} 105: 575-640. [arXiv:1310.5537]

\bibitem{GerardHHI} G\'erard C (2021) The Hartle-Hawking-Israel state on spacetimes with stationary bifurcate Killing horizons. {\it Reviews in Mathematical Physics} 33: 2150028.                                                                                  [arXiv:1806.07645]

\bibitem{PinSanVer}
Pinamonti N, Sanders K and Verch R (2019) Local incompatibility of the microlocal spectrum
condition with the KMS property along spacelike directions in quantum field theory on
curved spacetime.  {\it Lett.~Math.~Phys.} 109: 1735-1745. [arXiv:1806.02124]

\bibitem{KayLupo} Kay BS and Lupo U (2016) Non-existence of isometry-invariant Hadamard states for a Kruskal black hole in a box and for massless fields on 1+ 1 Minkowski spacetime with a uniformly accelerating mirror. {\it Classical and Quantum Gravity} 33: 215001. [arXiv: 1502.06582]

\bibitem{HawkingBHThermo} Hawking SW (1976) Black holes and thermodynamics.  {\it Phys.\ Rev.\ D} 13: 191-197.  

Quantization of fermions on Kerr space-time
\bibitem{CasalsEtAl} Casals M, Dolan SR, Nolan BC, Ottewill AC and Winstanley E {\it
Phys.~Rev.~D} 87: 064027. [arXiv:1207.7089]

\bibitem{DappEtAlUnruh} Dappiaggi C, Moretti V and Pinamonti N. (2011) Rigorous construction and Hadamard property of the Unruh state in Schwarzschild spacetime. {\it Advances in Theoretical and Mathematical Physics} 15: 355-447. [arXiv:0907.1034 ]

\bibitem{OriOttEtAl} (2022) Zilberman N, Casals M, Ori A and Ottewill AC (2022) Two-point function of a quantum scalar field in the interior region of a Kerr black hole.
{\it Phys.~Rev.~D} 106:125011. [arXiv:2203.07780]

\bibitem{GerHafWro} G\'erard C, H\"afner D and Wrochna M. (2020) The Unruh state for massless fermions on Kerr spacetime and its Hadamard property. arXiv:2008.10995.

\bibitem{PenBatelle} Penrose R (1968) Structure of space-time. {\it Battelle Rencontres. 1967 Lectures in Mathematics and Physics} deWitt C and Wheeler J (eds) (New York: Benjamin)

\bibitem{SimpsonPenrose} Simpson M and Penrose R (1973) Internal instability in a Reissner-Nordstr\"om black hole {International Journal of Theoretical Physics} 7: 183-187.

\bibitem{MisnerTaub} Misner CW and Taub AH (1969) A singularity-free empty universe.  {\it Soviet Physics JETP} 28: 122-133.

\bibitem{PenroseStrong} Penrose  R (1974) Gravitational collapse.  In: Dewitt-Morette C ed. {\it Gravitational radiation and gravitational collapse. International Astronomical Union Symposium No.~64.} pp. 82-91. (Dordrecht: Reidel) 

\bibitem{Landsman} Landsman K (2022) Penrose's 1965 singularity theorem: from geodesic incompleteness to cosmic censorship.  {\it General Relativity and Gravitation} 54: 115. [arXiv:2205.01680]

\bibitem{DafShlapRough} Dafermos M and Shlapentokh-Rothman Y (2018) Rough initial data and the strength of the blue-shift instability on cosmological black holes with $\Lambda > 0$. {\it Classical and Quantum Gravity} 35: 195010.  [arXiv:1805.08764]

\bibitem{ChrisDou} Christodoulou  D (2009) {\it The Formation of Black Holes in General Relativity (EMS Monographs in Mathematics)} (Z\"urich: European Mathematical Society)

\bibitem{LanirEtAl} Lanir A, Ori A, Zilberman N, Sela O, Maline A and Levi A (2019) Analysis of quantum effects inside spherical charged black holes. {\it Physical Review D} 99: 061502.
[arXiv:1811.03672]

\bibitem{OriOttPRL} Zilberman N, Casals M, Ori A and Ottewill AC (2022)  Quantum Fluxes at the Inner Horizon of a Spinning Black Hole.
{\it Phys.~Rev.~Lett.} 129: 261102 [arXiv:2203.08502].

\bibitem{BeniParti} Ju\'arez-Aubry BA (2015) Can a particle detector cross a Cauchy horizon?
{\it International Journal of Modern Physics D} 24:1542005. [arXiv:1502.02533]

\bibitem{BeniJorma} Ju\'arez-Aubry BA and Louko J (2022). Quantum kicks near a Cauchy horizon. {\it AVS Quantum Science}  4: 013201. [arXiv:2109.14601]

\bibitem{HintzVasy} Hintz P and Vasy A (2017) Analysis of linear waves near the Cauchy horizon of cosmological black holes. {\it Journal of Mathematical Physics} 58: 081509. [arXiv:1512.08004]

\bibitem{HWZ} Hollands S, Wald RM and Zahn J (2020) Quantum instability of the Cauchy horizon in Reissner-Nordstr\"om-deSitter spacetime. {\it Classical and Quantum Gravity} 37: 115009. [arXiv:1912.06047]

\bibitem{HKZ} Hollands S, Klein C and Zahn J (2020) Quantum stress tensor at the Cauchy horizon of the Reissner-Nordström-de Sitter spacetime. {\it Physical Review D} 102: 085004.
[arXiv:2006.10991]

\bibitem{MarUnr} Markovi\'c D and Unruh WG (1991) Vacuum for a massless scalar field outside a collapsing body in de Sitter space-time {\it Phys.~Rev.~D} 43:332.

\bibitem{KayNantes} Kay BS (2001) Application of linear hyperbolic PDE to linear quantum fields in curved space-times: Especially black holes, time machines and a new semilocal vacuum concept. {\it Journ\'ees \'Equations aux d\'eriv\'ees partielles, Nantes, 5 au 9 juin 2000, GDR 1151 (CNRS) pp. IX-1 to IX-19} (Also available at \url{http://www.math.sciences.univ-nantes.fr/edpa/2000/html}) [arXiv:gr-qc/0103056]

\bibitem{HolPhD} Hollands S (2000) Aspects of Quantum Field Theory on Curved Spacetime PhD Thesis University of York.

\bibitem{CKlein} Klein CKM (2023) Construction of the Unruh state for a real scalar field on the Kerr-de Sitter spacetime.  {\it Annales Henri Poincar\'e} 24: 2401-2442. 
[arXiv:2206.05073]

\bibitem{CramerKayTwo} Cramer CR and Kay BS (1998) Thermal and two-particle stress-energy must be ill defined on the two-dimensional Misner space chronology horizon. {\it Physical Review D} 57:1052. 
[arXiv:gr-qc/9708028]

\bibitem{Krasnikov} Krasnikov SV (1996) Quantum stability of the time machine. {\it Physical Review D} 54: 7322. [arXiv:gr-qc/9508038]

\bibitem{VisserReliab} Visser M (1997) The reliability horizon for semi-classical quantum gravity: Metric fluctuations are often more important than back-reaction. {\it Physics Letters B} 415: 8-14. [arXiv:gr-qc/9702041]

\bibitem{KayFlocality} Kay BS (1992) The principle of locality and quantum field theory on (non globally hyperbolic) curved spacetimes. {\it Reviews in Mathematical Physics} 4: 167-195.  
(Available at \url{https://www.researchgate.net/publication/234525000_The_Principle_of_Locality_and_Quantum_Field_Theory_on_non_Globally_Hyperbolic_Curved_Spacetimes})

\bibitem{Janssen} Janssen DW (2022) Quantum fields on semi-globally hyperbolic space-times. {\it Communications in Mathematical Physics} 391: 669-705. [arXiv:2111.01643]

\bibitem{MedaPinSiem} Meda P, Pinamonti N, Roncallo S and Zangh\`i N (2021) Evaporation of four-dimensional dynamical black holes sourced by the quantum trace anomaly. {\it Classical and Quantum Gravity} 38: 195022.
(See also Erratum (2022) {\it Classical and Quantum Gravity} 39: 059501.) [arXiv:2103.02057]

\bibitem{KayInfoLoss} Kay BS (2022) The black hole information loss puzzle, matter-gravity entanglement entropy and the second law.  arXiv:2206.07445v4.   (A shorter version of this appears will appear as:   Kay BS (2023) Matter-gravity entanglement entropy and the second law
for black holes.  {\it International Journal of Modern Physics D} (to appear) [arXiv:2305.11723])

\bibitem{HawkingInfoLoss} Hawking SW (1976) Breakdown of predictability in gravitational collapse. {\it Phys.~Rev.~D} 14: 2460.

\bibitem{UnruhWald} Unruh WG and Wald RM (2017) Information loss. {\it Reports on Progress in Physics} 80: 092002. [arXiv:1703.02140]

\bibitem{Maudlin} Maudlin T (2017) (Information) paradox lost. arXiv:1705.03541.

\bibitem{OkonSudarsky} Okon E and Sudarsky D (2015) The Black Hole Information Paradox and the Collapse of the Wave Function  {\it Found.~Phys.} 45:461-470 [arXiv:1710.01451]

\bibitem{BeniConj} Ju\'arez-Aubry BA (2023) Quantum strong cosmic censorship and black hole evaporation. arXiv:2305.01617v1.

\bibitem{HawkStew} Hawking SW and Stewart JM (1993). Naked and thunderbolt singularities in black hole evaporation. {\it Nuclear Physics B} 400: 393-415. [arXiv:hep-th/9207105]

\bibitem{AndEak} Anderson PR and Eaker W (1999) Analytic approximation and an improved method for computing the stress-energy of quantized scalar fields in Robertson-Walker spacetimes. {\it Physical Review D} 61: 024003. [arXiv:gr-qc/9906055]

\bibitem{LevOri} Levi A and Ori A (2016) Versatile method for renormalized stress-energy computation in black-hole spacetimes. {\it Physical Review Letters} 117: 231101. [arXiv:1608.03806]

\bibitem{TayBreeOtt} Taylor P, Breen C and Ottewill A (2022) Mode-sum prescription for the renormalized stress energy tensor on black hole spacetimes. {\it Physical Review D} 106: 065023. [arXiv: 2201.05174]

\bibitem{VerchSpinStat} Verch R (2001) A spin-statistics theorem for quantum fields on curved spacetime manifolds in a generally covariant framework. {\it Commun.\ Math.\ Phys.} 223: 261-288.

\bibitem{FewsterSpinStat} Fewster CJ (2016) On the spin-statistics connection in curved spacetimes. In: Finster F, Kleiner J, R\"oken C and Tolksdorf J (eds) {\it Quantum Mathematical Physics.} Cham: Birkh\"auser. [arXiv:1503.05797]

\bibitem{HollandsPCT} Hollands S (2004) A General PCT theorem for the operator product expansion in curved space-time.  {\it Commun.\ Math.\ Phys.} 244: 209-244. [arXiv:gr-qc/0212028]

\bibitem{StrohEtAl2002} Strohmaier A, Verch R and Wollenberg M (2002) Microlocal analysis of quantum fields on curved space-times: Analytic wave front sets and Reeh-Schlieder theorems. {\it J.\ Math.\ Phys.} 43: 5514. [arXiv:math-ph/0202003]

\bibitem{KoSanReehSchl} Sanders, K (2009) On the Reeh-Schlieder Property in Curved Spacetime. {\it Commun.\ Math.\ Phys.} 288: 271-285. [arXiv: 0801.4676]

\bibitem{StrohWitten} Strohmaier A and Witten E (2023) Analytic states in quantum field theory on curved spacetimes.  arXiv:2302.02709.

\bibitem{FewSplit1} Fewster CJ (2015) The split property for locally covariant quantum field theories in curved cpacetime. {\it Lett.~Math.~Phys.} 105: 1633-1661 [arXiv:1501.02682]

\bibitem{FewSplit2} Fewster CJ (2016) The split property for quantum field theories in flat and curved spacetimes. {\it Abh.~Math.~Semin.~Univ.~Hambg.} 86: 153-175 [arXiv:1601.06936]

\bibitem{FewstLectEnIn} Fewster CJ (2012) Lectures on energy inequalities. arXiv:1208.5399.

\bibitem{KonSanEnInequ} Kontou E-A, Sanders, K (2020) Energy conditions in general relativity and quantum field theory. [arXiv: 2003.01815]

\bibitem{WittenEnIneq} Witten E (2020) Remarks on energy inequalities.  YouTube video of conference talk  \url{https://www.youtube.com/watch?v=0Oh-Kmy-mx0}

\bibitem{HollYM} Hollands S (2008) Renormalized quantum Yang-Mills fields in curved spacetime. {\it Reviews in Mathematical Physics} 20: 1033-1172. [arXiv:0705.3340]

\bibitem{FredRejArt} Fredenhagen K and Rejzner K (2015) Perturbative construction of models of algebraic quantum field theory. In (Brunetti et al., 2015) (see `Further Reading').

\bibitem{RejznerBook} Rejzner K. (2016) Perturbative Algebraic Quantum Field Theory. {\it Math.~Phys.~Stud} Springer International Publishing.

\bibitem{RejznerEncyc} Brunetti, R., Fredenhagen, K. and Rejzner, K. (2025) Perturbative algebraic quantum field theory and beyond. Encyclopedia of Mathematical Physics 5, 464-476.

\bibitem{DutschBook} D\"utsch M (2019) {\it From Classical Field Theory to Perturbative Quantum Field Theory.} Cham: Birkh\"auser. 

\bibitem{FredRejBack} Fredenhagen K and Rejzner K (2012). Local covariance and background independence. Quantum Field Theory and Gravity: In:  Finster F, M\"uller O, Nardmann M, Tolksdorf J and Zeidler E (eds) (2012)
{\it Quantum Field Theory and Gravity, Conceptual and Mathematical Advances in the Search for a Unified Framework}  (Birkh\"auser, Springer-Basel) pp.~15-23. [arXiv:math-ph/1102.2376]

\bibitem{BrunFredRej} Brunetti R, Fredenhagen K and Rejzner K (2016) Quantum gravity from the point of view of locally covariant quantum field theory. {\it Communications in Mathematical Physics} 345: 741-779. [arXiv:1306.1058]

\bibitem{BrunFredHackPinRej} Brunetti R, Fredenhagen K, Hack TP, Pinamonti N and Rejzner K (2016) Cosmological perturbation theory and quantum gravity. Journal of High Energy Physics, 2016: 32. [arXiv:1605.02573]

\bibitem{Lima} Lima WCC (2021) Graviton backreaction on the local cosmological expansion in slow-roll inflation.  {\it Class.~Quantum~Grav.} 38: 135015 [arXiv:2007.04995]

\bibitem{FrobReinVerch} Fr\"ob MB, Rein C and Verch R (2022) Graviton corrections to the Newtonian potential using invariant observables. {\it Journal of High Energy Physics} 2022: 180. [arXiv:2109.09753]

\bibitem{BrunFredRejEff}  Brunetti R, Fredenhagen K and Rejzner K (2022) Locally covariant approach to effective quantum gravity. arXiv:2212.07800.

\bibitem{BJM} Barata JCA, J\"akel CD and Mund J (2023) The $P(\phi)_2$ model on de Sitter space. {\it Memoirs of the American Mathematical Society} 281: 1389.


\bibitem{Ford} L.H.~Ford (2005) Spacetime in semiclassical gravity. In Ashtekar A (ed) {\it 100 Years of Relativity -- Space-time Structure: Einstein and Beyond} (Singapore, World Scientific) [arXiv:gr-qc/0504096]

\bibitem{HuVerd} Hu BLB and Verdaguer E (2020) {\it Semiclassical and Stochastic Gravity: Quantum Field Effects on Curved Spacetime.} Cambridge, Cambridge University Press.

\bibitem{BeniCosConst} Ju\'arez-Aubry BA (2019) Semi-classical gravity in de Sitter spacetime and the cosmological constant. {\it Physics Letters B} 797: 134912. [arXiv:1903.03924]

\bibitem{BeniTonDan} Ju\'arez-Aubry BA, Miramontes T, Sudarsky D (2020) Semiclassical theories as initial value problems.  {\it J.~Math.~Phys.} 61: 032301 [arXiv:1907.09960]

\bibitem{GottSiem} Gottschalk H and Siemssen D. (2021) The cosmological semiclassical Einstein equation as an infinite-dimensional dynamical system.  {\it Annales Henri Poincar\'e} 22: 3915-3964. [arXiv:1809.03812]

\bibitem{MedPinSiem} Meda P, Pinamonti N and Siemssen D (2021) Existence and uniqueness of solutions of the semiclassical Einstein equation in cosmological models. {\it Annales Henri Poincar\'e} 22:3965-4015.  [arXiv:2007.14665]

\bibitem{GottRothSiem1} Gottschalk H, Rothe NR, Siemssen D (2022) Special cosmological models derived from the semiclassical Einstein equation on flat FLRW space-times.
{\it Classical and Quantum Gravity} 39: 125004 [arXiv:2112.15050] 

\bibitem{KoSanSemi} Sanders K (2022) Static symmetric solutions of the semi-classical Einstein-Klein-Gordon system.   {\it Annales Henri Poincar\'e} 23: 1321-1358. [arXiv:2007.14311]

\bibitem{BeniSemiConstr} Ju\'arez-Aubry BA (2022) Semiclassical gravity in static spacetimes as a constrained initial value problem. {\it Annales Henri Poincar\'e} 23: 1451-1487 [arXiv:2011.05947]

\bibitem{BeniModak} Ju\'arez-Aubry BA and Modak SK (2022) Semiclassical gravity with a conformally covariant field in globally hyperbolic spacetimes. {\it Journal of Mathematical Physics} 63: 092303. [arXiv:2110.01719]

\bibitem{MedPin} Meda P, Pinamonti N (2023) Linear stability of semiclassical theories of gravity. {\it Annales Henri Poincar\'e} 24: 1211-1243. [arXiv:2201.10288]

\bibitem{GottRothSiem2} Gottschalk H, Rothe NR, Siemssen D (2023) Cosmological de Sitter Solutions of the Semiclassical Einstein Equation. {\it Annales Henri Poincar\'e} 24: 2949–3029. [arXiv:2206.07774]


\end{thebibliography}
\end{document}